\documentclass[useAMS,usenatbib]{mn2e}
\usepackage{epsfig}
\usepackage{epsf,rotating}
\usepackage{amsmath}
\usepackage{subfigure}
\usepackage{lscape}
\usepackage{comment}

\usepackage{natbib}

\newcommand{\HI}{H{\sc I}}

\newcommand{\kms}{\;{\rm km}\,{\rm s}^{-1}}

\newcommand{\gad}{{\sc Gadget-2}}
\newcommand{\autovp}{{\sc AutoVP}}
\newcommand{\ion}[2]{\hbox{#1\,{\sc #2}}}
\newcommand{\vw}{v_{\rm w}}
\newcommand{\cm}{\,{\rm cm}}
\newcommand{\apj}{ApJ}
\newcommand{\mnras}{MNRAS}
\newcommand{\apjs}{ApJS}

\newcommand{\msolar}{\;{\rm M}_{\odot} }

\newcommand{\comm}[1]{}

\begin{document}

\title[Cycling in Low-Z CGM]{Baryon Cycling in the Low-Redshift Circumgalactic Medium: A Comparison of Simulations to the COS-Halos Survey}  

\author[Ford et al.]{
\parbox[t]{\textwidth}{\vspace{-1cm}
Amanda Brady Ford$^1$, Jessica K. Werk$^2$, Romeel Dav\'{e}$^{3,4,5}$,
Jason Tumlinson$^6$, Rongmon Bordoloi $^6$, Neal Katz$^7$, Juna A. Kollmeier$^8$, Benjamin D. 
Oppenheimer$^9$, Molly S. Peeples$^6$, Jason X. Prochaska$^2$, David H. Weinberg$^{10}$}
\\\\$^1$ Max Planck Institute for Astrophysics, D-85741, Garching, Germany 
\\$^2$ UCO/Lick, University of California Santa Cruz, Santa Cruz, CA, 95064
\\$^3$ University of the Western Cape, Bellville, Cape Town 7535, South Africa
\\$^4$ South African Astronomical Observatories, Observatory, Cape Town 7925, South Africa
\\$^5$ African Institute for Mathematical Sciences, Muizenberg, Cape Town 7945, South Africa
\\$^6$ Space Telescope Science Institute, 3700 San Martin Dr, Baltimore, MD 21218
\\$^7$ Astronomy Department, University of Massachusetts, Amherst, MA 01003, USA
\\$^8$ Observatories of the Carnegie Institution of Washington, Pasadena, CA 91101, USA
\\$^9$ CASA, Department of Astrophysical and Planetary Sciences, University of Colorado, Boulder, CO 80309, USA
\\$^{10}$ Astronomy Department and CCAPP, Ohio State University, Columbus, OH 
43210, USA
}

\pubyear{2015}
\label{firstpage}

\maketitle
 \begin{abstract}
We analyze the low-redshift (z$\approx$0.2) circumgalactic medium
by comparing absorption-line data from the COS-Halos Survey to
absorption around a matched galaxy sample from two cosmological
hydrodynamic simulations. The models include different prescriptions
for galactic outflows, namely hybrid energy/momentum driven wind
(ezw), and constant winds (cw).  We extract for comparison direct
observables including equivalent widths, covering factors, ion
ratios, and kinematics.  Both wind models are generally in good
agreement with these observations for \ion{H}{i} and certain low
ionization metal lines, but show poorer agreement with higher
ionization metal lines including \ion{Si}{iii} and \ion{O}{vi} that
are well-observed by COS-Halos.  These discrepancies suggest that
both wind models predict too much cool, metal-enriched gas and not
enough hot gas, and/or that the metals are not sufficiently well-mixed.
This may reflect our model assumption of ejecting outflows as cool
and unmixing gas.  Our ezw simulation includes a heuristic prescription
to quench massive galaxies by super-heating its ISM gas, which we
show yields sufficient low ionisation absorption to be broadly
consistent with observations, but also substantial \ion{O}{vi}
absorption that is inconsistent with data, suggesting that gas
around quenched galaxies in the real Universe does not cool.  At
impact parameters of $\la 50$~kpc, recycling winds dominate the
absorption of low ions and even \ion{H}{i}, while \ion{O}{vi} almost
always arises from metals ejected longer than 1~Gyr ago.  The
similarity between the wind models is surprising, since we show
that they differ substantially in their predicted amount and phase
distribution of halo gas. We show that this similarity owes mainly
to our comparison here at fixed stellar mass rather than at fixed
halo mass in our previous works, which suggests that CGM properties
are more closely tied to the stellar mass of galaxies rather than
halo mass.
\end{abstract}

\section{Introduction}

Studying the circumgalactic medium (CGM) is an emerging frontier
of galaxy evolution that connects galaxies to the intergalactic
medium. In this tenuous multi-phase gas that surrounds galaxies lie
clues to the processes that drive galaxy evolution such as gaseous
inflows and outflows~\citep[e.g.][]{ker05,dek09,dav12,lil13}.  Recent
observations with {\it Hubble's} Cosmic Origin Spectrograph (COS)
have characterised the CGM in unprecedented detail. They have shown that there
is probably more oxygen in the halos around star-forming galaxies
than within them \citep{tum11}, that the CGM may contain the missing
baryons and missing metals within galaxy halos \citep{pee14,shu14,wer14},
and that star-forming and passive galaxies have different \ion{O}{vi}
signatures \citep{tum11} but similar cool gas content~\citep{tho12,wer14}.  Recent work by \citet{sto13} and
\citet{tri12} has also used COS to probe the CGM out to several
hundred kiloparsecs, and observations by \citet{pro11} and \citet{rud13}
have established a characteristic extent of the metal-enriched CGM
to be roughly $\sim$ 300~proper kpc, regardless of redshift.  In
addition to these observations that characterise the CGM, there is
a wealth of observational data detecting outflows at a variety of
redshifts \citep{mar05, rup05, tre07,wei09, rub12, pet01,ste01,vei05},
indicating that most if not all galaxies drive winds at some point
in their evolution, presumably enriching the CGM and IGM with mass
and metals in the process.

The physical mechanisms that drive these outflows are not well
characterised.  While observations can now probe individual sight
lines around individual galaxies, there is not yet a robust
understanding of the metallicity, covering factor, and thermodynamic
phase of the outflowing material. The specifics of how these
quantities vary with the galaxy's mass, star formation rate,
environment, and redshift are poorly characterised.  The fate of the
ejected material after it leaves the galaxy is even less well
constrained. Simulations suggest the presence of ``halo fountains",
where baryons are ejected from the galactic disk only to rain back
down again at later times~\citep{opp08}. The contribution of this so-called recycled accretion
relative to pristine inflows from the IGM depends at minimum on
galaxy mass~\citep{opp10} and impact parameter \citep{for13b}, and
plausibly also on star formation rate (SFR), SFR surface density, and
environment.  Simulations by \citet{opp12} suggested that
the IGM is enriched ``outside-in", where winds propagate further
from galaxies at earlier epochs and metals re-collect towards halos
at late times, but this conclusion may depend on the poorly-understood
mechanisms by which winds are launched.

Hydrodynamic simulations that directly model inflows and outflows
provide a crucial complement to emerging CGM observations by plausibly
elucidating how the physical conditions and dynamical state of the
CGM gas can be manifest in absorption line data. \citet{for13a}
showed that metal ions with low ionisation potential (e.g.,
\ion{Mg}{II}) trace dense gas close to galaxies, while metals with
high ionisation potentials (e.g., \ion{O}{vi}) trace more diffuse
gas. This work also showed how different prescriptions for outflows
can result in significant differences in the temperature, density, and
metallicity of the CGM gas. \citet{for13b} followed and showed, in
detail, how the baryon cycle operates in a plausible wind model.
At low redshift, inflowing material is dominated by recycled gas
that was previously ejected in an outflow; hence inflows are enriched
and can be traced by metal lines. In general, low metal ions like
\ion{Mg}{ii} trace this recycled accretion, while \ion{O}{vi} traces
hotter, less dense material built up by prior epochs of outflows.

While these simulations have provided some general guidelines for
interpreting CGM observations, they have yet to be tested in detail
against existing CGM data.  The best existing absorption line data
set that targets a range of impact parameters close to galaxies in
the CGM is currently from the COS-Halos
survey~\citep{tum11,tum13,wer12,wer13}.  This survey targets 39
quasars around 44 $0.1L^* - 3~L^*$ galaxies with a carefully-selected
sampling of colour and stellar mass, and constitutes the first galaxy-selected
absorption line survey of low-redshift CGM gas.  These galaxies
have stellar masses ranging from $10^{9.5-11.5}$, with spectroscopic
redshifts $z_{phot} \approx 0.14-0.34$, and are located $15-160$~kpc
(projected) from a background UV-bright quasar.  While COS-Halos
covers a wide range of ions, for this work we focus on \ion{H}{i}1216,
\ion{Si}{ii}1260, \ion{C}{ii}1335, \ion{Si}{iii}1206, \ion{Si}{iv}1394,
\ion{C}{iii}977, and \ion{O}{vi}, since they span a representative range of
ionisation energies (see Table \ref{table:thresh}).  Typical COS-Halos
detection thresholds for the various ions are given in Table
\ref{table:thresh}, and are generally around rest equivalent widths
of 0.1\AA.  The COS-Halos sight lines are supplemented by Keck/HIRES
echelle spectra for 35 quasars to include coverage for \ion{Mg}{ii}
(2796, 2803\AA).  These observations represent the largest sample of CGM
absorption line data at these impact parameters, and hence provide
a stringent test for models of CGM gas and their relationship to
galaxies. 

In this paper we build on our previous works using hydrodynamic
simulations to (i) determine if our hydrodynamic simulations
quantitatively reproduce the COS-Halos absorption line observations;
(ii) test whether such observations can distinguish between plausible
but substantively different models for galactic outflows; and (iii)
constrain the physical and dynamical state of CGM gas around $\sim
L^*$ galaxies as probed by COS-Halos data.  As observations improve with
ongoing surveys such as COS-Dwarfs \citep{bor14}, the COS-GTO survey
\citep{dan10, dan11, dan14, sto13}, and other efforts \citep[e.g.][]{hec13} to
study the CGM of selected galaxy samples, this work aims to serve
as a template for how CGM absorption line observations can constrain
the physical processes driving the baryon cycle.

Our paper is organised as follows: in \S 2 we describe our simulations,
in \S 3  we compare to direct observables, in \S 4 we describe variations
with wind model including variations in baryon cycling, and in \S 5 we
summarise our conclusions.

\section{Simulations and artificial spectra}

\subsection{The code and input physics}
\label{sec:code}

For this paper we use our modified version of the N-body+entropy-conserving
smooth particle hydrodynamic (EC-SPH) code \gad~\citep{spr05}, fully
described in \citet{opp08}. Our main simulation is identical to
that used in \cite{dav13,for13b}, and further details can be found
there. Briefly, we assume  a WMAP-9 concordant $\Lambda$CDM cosmology
\citep{hin09}: $\Omega_M = 0.28$, $\Omega_\Lambda = 0.72$, $h={\rm
H}_{0}/(100 \kms Mpc^{-1})=0.7$, a primordial power spectrum index
$n$ =0.96, an amplitude of the mass fluctuations scaled to
$\sigma_8$=0.82, and   $\Omega_b = 0.046$. The volume is 32${h}^{-1}$
Mpc on a side with ${512}^{3}$ gas particles and ${512}^{3}$ dark
matter particles. This results in a gas particle mass of
$4.5\times{10}^{6}M_\odot$ and
a dark matter particle mass of $2.3\times{10}^{7}M_\odot$, allowing us to 
resolve galaxies down to stellar masses approaching $10^8 M_\odot$. The main
simulation output we use for this work is at $z=0.25$, which is
the closest snapshot we have to the typical redshift of COS-Halos.

This simulation includes galactic outflows, which are implemented using a Monte Carlo approach and tied to the
star formation rate (SFR) via $\dot{M}_{wind} = \eta\times$SFR,
where $\eta$ is the mass loading factor. For this work we select
two wind models to study in detail. The first is our hybrid
energy/momentum driven winds or ``ezw" model. This model is our
favoured wind model because it most accurately represents the
relevant small-scale physics associated with the transition from
momentum-driven winds to energy-driven winds at low masses~\citep{hop13,mur15},
as well as reproducing observations of the stellar and \ion{H}{i}
mass function \citep{dav13}.  In the ezw model, the wind speed $\vw$
and mass loading factor $\eta$ depend on the galaxy velocity
dispersion $\sigma$:

\begin{eqnarray}
\vw &=& 4.29\sigma \sqrt{{\rm f}_{L}-1} + 2.9\sigma \\
\eta&=&\sigma_{o}/\sigma, {\rm\ if\ } \sigma > 75\kms\\
\eta&=&(\sigma_{o}/\sigma)^{2}, {\rm \ if\ } \sigma < 75\kms
\end{eqnarray}

Here, ${\rm f}_{L}$=[1.05,2]  is the luminosity in units of the
Eddington luminosity required to expel gas from a galaxy potential,
$\sigma_{0} = 150 \kms$, and $\sigma$ is the galaxy's internal
velocity dispersion, broadly constrained to match IGM enrichment
at high redshift \citep{opp08}. Particles are ejected individually,
with velocities chosen to lie within the range determined by this
range of ${\rm f}_{L}$. The range of ${\rm f}_{L}$ values is taken
from observations of local starburst galaxies by \citep{rup05}, and
is also consistent with observations by \citep{mar05} that find
${\rm f}_{L}\sim 2$.  In this model, as in its earlier cousin
vzw which did not include the steeper $\eta$ scaling at low $\sigma$,
we add an extra kick of $2.9\sigma$ to simulate continuous pumping
of gas from radiation pressure, as described in \citet{opp08}
\footnote{Our previous paper listed an incorrect formula for $\vw$;
the one shown here is verified to be the one used in our code.}.
This simulation also includes a prescription for quenching massive
galaxies, which involves super-heating particles with increasing
probability above a threshold quenching stellar mass as described
in \citet{dav13}.

Our second wind model is a constant wind (``cw") model, where the
mass loading factors and wind speeds are the same for each galaxy,
regardless of mass. In the cw model, $\eta=2$ and $v_w=680\kms$.
The latter is similar to that used in the Overwhelmingly Large
Simulations (OWLS) reference model of \cite {sch10}. The cw simulation
used here has the same physics described in \citet{for13a}, including
\cite{wie09a} metal-line cooling, but now with a box size and resolution
to match our main ezw simulation used here, namely 
32${h}^{-1}$ Mpc and $2\times {512}^{3}$ particles. This allows for consistency
with the ``ezw" model above, and easily resolves galaxies
similar in size to the smallest ones in the COS-Halos survey. 

The constant wind model does less well reproducing the observed
measurements of \ion{C}{iv} at high redshift~\citep{opp06}, the
mass-metallicity relation~\citep{fin08,dav11b}, and the evolution
of stellar growth in galaxies~\citep{dav11a}.  Nonetheless, it
viably reproduces the evolution of global star formation~\citep{spr03}.
However, it has not yet been tested against the COS-Halos survey
at low-z.  We include this wind model because it serves as a useful
contrast to the variable mass loading and wind speed wind model
ezw.  Through this comparison we aim to comment on not just {\it
whether} winds are necessary to match CGM observations, but also
{\it what type} of winds are necessary. We can also gain insight
into how baryons cycle in the different wind models, how they affect
the gas around galaxies, and how such processes are manifested in
CGM observations.

\subsection{Generating spectra with specexbin}

Once we run our simulations, we use {\sc Specexbin} to calculate
physical properties of the gas along lines of sight~\citep{opp06}.
Briefly, {\sc Specexbin} averages physical properties of the gas
along a given sight line and then uses look-up tables calculated
with {\sc CLOUDY} \citep[{}][version 08.00]{fer98}  to find the
ionisation fraction for the relevant ionic species. We use the same
version of {\sc Specexbin} as in \cite{for13a,for13b}. See Figure
1 of \citet{for13a} for an example of a simulated spectrum. This
version of {\sc Specexbin} includes a prescription for self-shielding
from the ionisation background, which uses a density threshold of
${\rm n}_{\rm H}=0.01 {\rm cm}^{-3}$. Above this density threshold
we assume that all the hydrogen is fully neutral, i.e. it is all
\ion{H}{i}, and that all the magnesium is in \ion{Mg}{ii}. An \HI\
column density above $\sim 10^{17}\cm^{-2}$ shields against most
of the 15eV photons that could ionize \ion{Mg}{ii} to \ion{Mg}{iii},
but it does not shield the 7eV photons that ionize \ion{Mg}{i} to
\ion{Mg}{ii}.

We choose a signal-to-noise (S/N) value of 30, for comparison with
our earlier work \citep{for13a,for13b}. This is higher than the S/N
value of the observations, which averages roughly 12. However, since
we will use the same equivalent width detection thresholds as
observations, as listed in Table \ref{table:thresh}, our results
are not very sensitive to our exact choice of S/N.  Once the model
lines of sight are selected (see the next section) and the artificial
spectra are generated using {\sc Specexbin}, then we use the Voigt
profile fitter \autovp\ \citep{dav97} to identify the absorption
lines (see \cite{for13a} for an example) and to fit column densities
and equivalent widths.

We choose a large suite of ions to examine for this work to get the
fullest possible information on the physical conditions of the CGM.
As found in earlier work \citep{wer12, for13a, for13b}, a range of
ionisation potentials are necessary to probe different phases of
the CGM. Generally, low ionisation potential metal lines probe high
density, low temperature regions close to galaxies, while high
ionisation potential metal lines probe warmer, more diffuse gas
that can be detected further from galaxies. In addition to
\ion{H}{i}1216, we generate spectra for \ion{Mg}{ii}2796,
\ion{Si}{ii}1260, \ion{C}{ii}1335, \ion{Si}{iii}1206,  \ion{Si}{iv}1394,
\ion{C}{iii}977, and \ion{O}{vi}1032. These metal lines are all
detected in COS-Halos, and probe a range of ionisation energies
(see Table \ref{table:thresh}).

\begin{table}
\caption{Ion Properties}
\begin{tabular}{lccc}
\hline
Ion & Ionisation Energy (eV) &
Detection Threshold (\AA)
\\
\hline
\multicolumn{3}{c}{}\\
\HI\ & 13.6 & 0.2 \\
\ion{Mg}{ii} & 15.04 & 0.1 \\
\ion{Si}{ii} & 16.35 & 0.15 \\
\ion{C}{ii}\ & 24.38 & 0.08 \\
\ion{Si}{iii} & 33.49 & 0.1 \\
\ion{Si}{iv} & 45.14 & 0.1 \\
\ion{C}{iii} & 47.89 & 0.25 \\
\ion{O}{vi} & 138.1 & 0.1 \\
\hline
\end{tabular}
\label{table:thresh}
\end{table}

\subsection{Line of sight selection}
\label{sec:LOS}

\begin{figure}
 \subfigure{\setlength{\epsfxsize}{0.50\textwidth}\epsfbox{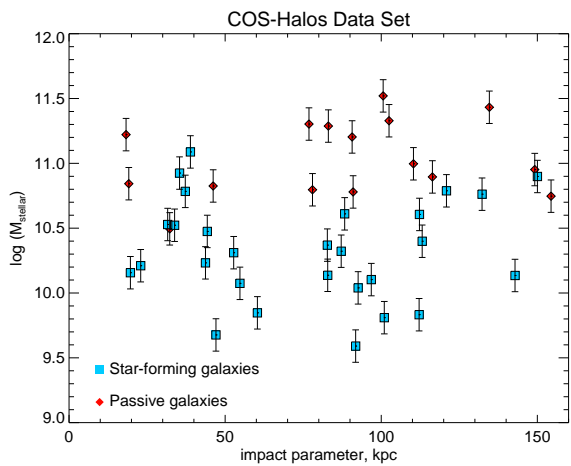}} %smrho_jess.pro 
\caption{Stellar mass vs. impact parameter for the observed COS-Halos survey. Star forming galaxies are separated from passive galaxies at ${\rm sSFR}=10^{-11}~M_\odot/yr$. Range bars show the mass range of galaxies in the simulations deemed comparable to those observed and do not represent errors.}
 \label{LOS} 
 \end{figure}

We choose our lines of sight (LOS) in our simulations to best mimic
the LOS in the COS-Halos survey. In COS-Halos, galaxies were selected
first and the impact parameter to the galaxy is 
the projected distance to the observed quasar. As a result, each
COS-Halos LOS corresponds to a specific impact parameter $b$ around
a galaxy of a known stellar mass $M_*$, as shown in Figure \ref{LOS}.
Moreover, COS-Halos galaxies were generally selected to be isolated
systems (preferring galaxies without photometric coincidences within 1~Mpc, with some exceptions), and hence are likely to be central galaxies in their
halos~\citep[Section 2.5]{tum13}. 

\begin{figure}
 \subfigure{\setlength{\epsfxsize}{0.55\textwidth}\epsfbox{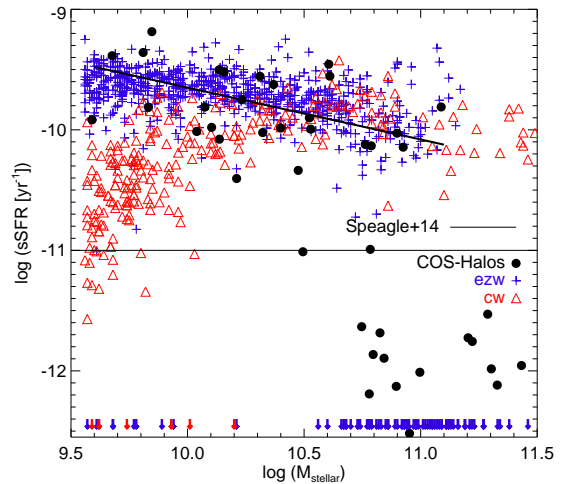}}
\caption{sSFR rate vs. stellar mass for the galaxies in the ezw (blue crosses) and cw (red triangles) models that match the COS-Halos LOS (black circles). Galaxies with sSFR=0 are plotted as downward facing arrows near the bottom. Overplotted in thick black is the median trend compiled from observations by \citep{spe14}. The thin black horizontal line represents the dividing line between passive and star-forming galaxies.} 
\label{sfr}
\end{figure}

To mimic this process in the simulations, we select all the central
galaxies in our simulations that are within $\pm$0.125 dex in stellar
mass of the COS-Halos survey galaxies, and we generate four LOS at
the particular impact parameter for the selected COS-Halos galaxy,
x+$b$, x-$b$, y+$b$, and y-$b$. In this way we produce a suite of model LOS for
each COS-Halos LOS that matches its unique ($b,M_*$) value.  For example,
for the COS-Halos LOS corresponding to $M_{*}=10^{10.2} M_\odot$ and
$b=18.26~{\rm kpc}$, we select all galaxies in our simulation with
$M_{*}=10^{10.075-10.325}$, and produce four LOS around each at
$b=\pm 18.26$~kpc in $x$ and $y$.  In the end, for each of the 44
COS-Halos ($b,M_*$) LOS, we have between 50-200 simulated LOS.
The smaller mass galaxies tend to have larger numbers of LOS since
there are more galaxies within $\pm$0.125 dex in $M_*$.  

For consistency with the COS-Halos analysis, we consider all
absorption components within $\pm$ 600 $\kms$ to be associated with
the galaxy, and sum them for a total equivalent width, unless
otherwise stated. We note that this differs from our approach in
earlier work \citep{for13a} where we only considered components
within $\pm$ 300 $\kms$ to be associated with a galaxy. However,
as shown in Figures 8 and 15 of that work, the absorption drops off
substantially after $\pm$ 300 $\kms$, so there is little difference
between these two limits. This same drop-off in absorption after
$\pm$ 300 $\kms$ is seen in the COS-Halos data themselves \citep{tum13}.

We note that our simulation naturally produces many satellite galaxies around
the chosen central galaxy, so in some cases our LOS will impact
closer to satellites.  This will presumably contribute to the
variance of absorption properties around a particular galaxy.  In
this paper, we will focus on mean and median absorption statistics
without examining the origin of the scatter, which we leave for
future work.

We also subdivide our sample into passive and star-forming galaxies
(SFGs), using a specific star formation rate (sSFR$\equiv$SFR/$M_*$)
criterion of $10^{-11} M_\odot/$yr, as done for COS-Halos (see
Figure~\ref{LOS}).  Figure \ref{sfr} shows sSFR vs. $M_*$ for ezw
(blue crosses) and cw (red triangles) models, only showing those
galaxies selected to mimic the COS-Halos LOS.  We show as the black
line the median trend from a compilation of observations of
star-forming galaxies by \citet{spe14}. The observed COS-Halos
galaxies are generally actively star forming for the lower-mass
galaxies, but are more likely to be passive with increasing stellar
mass. Many of the massive galaxies in ezw have zero SFR, and those
quenched galaxies are plotted at the bottom as downward facing
arrows.  Of the 48 $M_{*}> 10^{11} M_\odot$ galaxies in the ezw
sample, all but 9 are quenched, with sSFR=0.  Unlike the ezw model,
cw does not include a quenching prescription, so produces no quenched
galaxies at high masses.  The simulated galaxies broadly cover
the range of sSFRs seen in COS-Halos, albeit that our quenched
galaxies have zero SFR by construction.  

Figure \ref{sfr} only shows galaxies identified as central galaxies
within their halos, but at low masses, many of the quenched galaxies
shown here are likely former satellites that have been ejected from
halos~\citep{gab14}; our quenching prescription is not applied to
such small systems, but they have been quenched as a satellite
likely by gas stripping processes~\citep[e.g.][]{raf14}. Such
satellite quenching processes are self-consistently handled in the
simulations, albeit the version of SPH we use here does not necessarily
handle them optimally~\citep[e.g.][]{age07,hop13}.  Importantly,
we note that the ezw model better matches the observed distribution
of sSFR vs. $M_*$ at low masses.  This is one reason why we
favour our ezw model relative to cw, in addition to those arguments
discussed in \S~\ref{sec:code}. We also favour the ezw model because
it more accurately encapsulates the scalings for outflows predicted
from recent high-resolution individual galaxy simulations~\citep{mur15},
and matches many other observations better than cw~\citep{dav13}.

\section{Comparison to COS-Halos Observables}

We now show a direct comparison between our simulations and the
COS-Halos survey. This work represents the first direct comparison
of our simulations against observations of absorption lines in
targeted LOS at low redshift.  Each of the following subsections
represents a different way of comparing models to observations; all must
agree before one can claim agreement. We begin by focusing on the
strength of {\it detected} lines, to see if the models match the
observed sample.  Then we examine the total absorption along all
LOS, as well as the incidence rate of all our ions along all LOS.
We then provide a preliminary glimpse into how ion ratios can
constrain models, and examine the kinematics of absorbers relative
to their host galaxies.  Matches and mismatches between all these
statistics provide insights into exactly how our simulations are
succeeding and failing in reproducing the COS-Halos data, and 
they points toward required improvements in CGM modelling.

\subsection{Equivalent widths vs. impact parameter}

\begin{figure*}
 \centering
% \includegraphics{ew.jess.oct27.ps}
% \makebox[\textwidth]{\includegraphics[width=.9\paperwidth]{ew.jess.oct27.ps}}
\subfigure{\setlength{\epsfxsize}{1.1\textwidth}\epsfbox{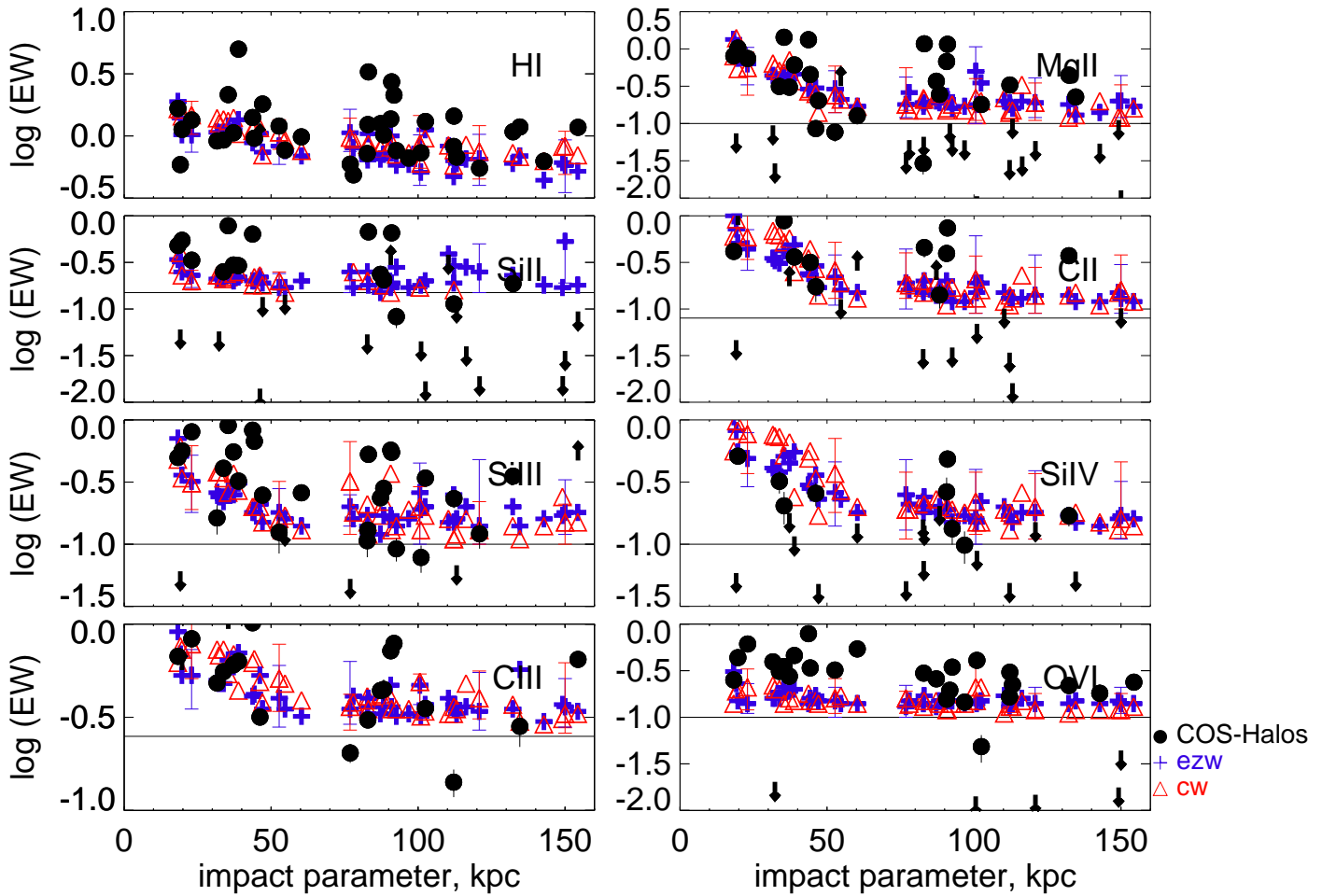}}
\caption{Equivalent width (\AA) vs. impact parameter for observed points (black --- circles for measured values, up and down facing arrows for lower and upper limits respectively); as well as ezw (blue crosses) and cw (red triangles) models. Model points are the median values over LOS with detections above the thresholds given in Table \ref{table:thresh}, range bars show the 16-84\% values of those LOS with detections. Ions are ordered according to increasing ionisation energy (right to left, top to bottom). Model points do not smoothly decrease, owing to conflation between stellar mass and impact parameter (see Figure \ref{LOS}). Errors on observed points are given for detections only (not limits) and represent the 2$\sigma$ limits. }
\label{ew}
\end{figure*}

The most direct absorption line observable is the equivalent width.
Figure~\ref{ew} shows the rest equivalent width versus impact
parameter for the COS-Halos survey compared to our simulations, for
our set of probed ions. For this figure we focus on {\it detected} lines
only, and ask if models reproduce the same absorption strength as
the detected observations. Note that the detection thresholds for
each ion are given in Table \ref{table:thresh}, and shown as solid
black horizontal lines to guide the eye. We note that for the
observations, the detection limits quoted are a rough guide, as
they can vary vary for individual LOS. For the larger sample of LOS
in the models, however, we use the detection limits quoted in
Table \ref{table:thresh} for all model LOS.  Observed points below the detection limits are included in this figure for completeness, however in this work we only
compare observed and model points {\it above the detection limits}.

Here we have placed the ions in order of increasing ionisation
potential from low to high. The black points  are from the COS-Halos
survey. The circles represent detected values while the up and down
facing arrows are lower and upper limits, respectively.  Error bars
are included in black for detected values only, errors on limits
are not included. In many cases the errors are smaller than the size of
the symbols.

The blue crosses and red triangles show the results for the ezw and
cw models, respectively. These points represent the median equivalent
width of all LOS {\it with detections above the threshold} listed in Table \ref{table:thresh} at a given
impact parameter. The bars on the model points span the 16-84\%
range of EW values at that impact parameter, to illustrate the
scatter in EW at a given impact parameter. To avoid clutter, we
only show the model range bars for points $\approx$ every 25~kpc. The impact of non-detections will be better illustrated by the statistics shown in Figures \ref{dedz} and \ref{cf}.

Both models and observations show a weak trend of decreasing equivalent widths
with impact parameter.  As discussed in \cite{for13a}, the
basic driver of this is that gas is denser near galaxies.  However,
the trend is weaker for higher ions, and is essentially absent for
\ion{O}{vi}, because the density gradient is counteracted by the
tendency for high ions to have increasing ionisation fractions in
lower-density gas.  In \cite{for13a} we found that the ion strength
dropped with both halo mass (and thus stellar mass) and
impact parameter.  But it is important to note that Figure \ref{ew}
conflates these two trends, owing to the way that we have done our LOS
selection (\S \ref{sec:LOS}), by matching the observed galaxy sample
in {\it both} stellar mass and impact parameter.  As shown in Figure
\ref{LOS}, LOS at either low or high impact parameters can be from
either low or high mass galaxies. As such, the model points do not
decrease as strongly or smoothly with impact parameter, as was seen
in \cite{for13a} where we selected narrow bins in halo mass.
Additionally, this degeneracy washes away some trends with ionisation
potential: the equivalent widths of the low ionisation species do
not drop off as steeply as found in \citet{for13a,for13b}.  This
will similarly impact Figure~\ref{dedz}.  The scatter owing to the
conflation of these trends is exacerbated for the observed points,
since the simulation points are averages from a larger sample.

For \ion{H}{i}, both simulations are in general agreement with the
observations across the 0-150~kpc range, showing strong absorption in every
line of sight. The typical EW is much higher than seen
in random lines of sight through the IGM~\citep{dav10,for13a},
showing that there is substantial cool gas within these halos.
There is little sensitivity to winds, at least as can be discriminated
from the COS-Halos survey, which is reminiscent of the lack of
sensitivity to winds in the Ly$\alpha$ forest~\citep{dav10}.  There
are, however, a few very strong absorbers with equivalent widths
above 2\AA\ that lie above all the simulation median points.  These
potentially provide an interesting challenge to models, hinting
at a larger covering fraction of very strong \ion{H}{i} than
predicted.  However, current uncertainties prevent any robust
constraints emerging from these few systems.

The metal lines provide a more direct test of pollution from galactic
outflows.  For the low and mid ions, when detections are made, they
are generally in the EW range expected from the models, with a
few observed points being much higher than the model points (as with
\HI).  \ion{O}{vi} is also commonly detected, but the typical
equivalent widths are generally $\times 2-3$ higher than predicted;
we discuss this below.  Nonetheless, the broad agreement is a
non-trivial success of these galactic outflow models in enriching
the CGM to roughly the observed levels.  Though we do not show it, a
model without winds dramatically fails this basic test since it
does not yield appreciable metal absorption beyond the smallest impact parameters,
as discussed in \citet{for13a}.

For the low ions (\ion{Mg}{ii}, \ion{Si}{ii}, \ion{C}{ii}, and
\ion{Si}{iii}), the model points tend to lie generally in the middle
of the observed points. In many cases the scatter in the observed points is larger than the 16-84\% range bars that the models would predict. This could reflect variations
in the ionisation conditions or metallicities that are not properly
captured within our simulations.  Recall that we assume a
spatially-uniform metagalactic ionising background with no local
contribution from the nearby galaxy and a simple density criterion
for self-shielding. These assumptions may need to be relaxed if we
wish to capture detailed variations properly.  Doing so would require
full radiative transfer simulations with much higher resolution to
capture small, dense condensations.  As discussed in \citet{cri15},
if such clouds are indeed the dominant contribution to low-ionisation
CGM absorption, they are far from being resolved in any current
cosmologically-based simulation.

The mid and high ions tend to show a smaller observed scatter than
the low ions, more comparable to that seen in the simulations.
Since low ions tend to arise closer to galaxies and in denser
gas~\citep{for13a}, this suggests that locally-varying density and
ionisation conditions, and not locally-varying metallicity, is
responsible for the larger scatter in the low ions.  Alternatively,
local variations in metallicity could be much stronger closer to
galaxies, which may be reasonable if the absorbing gas arises from
recent outflows.

Returning to the discrepancy for \ion{O}{vi}, this is tricky to
interpret because \ion{O}{vi} can arise in both photo-ionised and
collisionally-ionised gas~\citep{opp09}.  As shown in \citet[Figure
6]{for13a}, the median temperature for \ion{O}{vi} absorption is
$\sim 10^{4.5}$K, which is dominated by photo-ionisation, and indeed
the phase space diagram in their Figure~5 does not suggest a distinct
peak of collisionally-ionised \ion{O}{VI} absorption at $T\sim
10^{5.5}$K.  One way to explain the predicted deficit is thus to
invoke a more significant hot halo of gas.  However, since the
typical virial temperature of a COS-Halos galaxy is a few times
$10^6$K, this means that such collisionally-ionised \ion{O}{VI}
would have to be a transient phase through which the gas is cooling.
Modelling such non-equilibrium processes is a substantial challenge
for simulations, and may need to be improved if more robust and
accurate predictions are desired.  Alternatively, a significantly
higher photo-ionising background around 114 eV than what we have
assumed could also result in substantially more \ion{O}{vi}; the
metagalactic flux at these energies is currently poorly known.

The details of the wind implementation can strongly impact the phase of CGM
gas.  For instance, the models of \citet{hum13} evolved using the adaptive
mesh refinement hydrodynamic code {\sc Enzo} show substantially
hotter CGM gas, since their models drive winds by super-heating the
ISM.  They tended to find that the resulting CGM was too hot, and
thus also a poor match to \ion{O}{vi} observations, at least for the most
plausible assumptions about the energetic input into winds.  This
highlights that the proper treatment of wind propagation, heating,
and mixing is a challenging problem in current simulations, and
shows that detailed comparisons to observations can provide interesting
constraints on these processes.

In light of this, it is somewhat surprising that the two wind models
we consider here are in such good agreement with each other, and
that these COS-Halos observations cannot discriminate between them.
\cite{for13a} showed large differences in absorption line profiles
in every ion between a cw and a momentum-driven wind model that is
similar to ezw.  Moreover, ezw and cw are well-discriminated by
other diagnostics generally related to galaxy properties \citep{dav13}.
The differences between wind models will be slightly more prominent
in other CGM statistics that we explore later, but for the equivalent
widths they are small.  We will discuss potential reasons for this
in \S4. 

\subsection{Equivalent width per unit redshift}

In the previous section we compared individual absorber equivalent
widths between the models and the observations where detections occurred, and
found broad agreement in some ions but notable differences in others.
That comparison, however, does not characterise the {\it total}
absorption strength; for instance, it may be that absorbers are far
more or less common in the models compared to the observations, so that
while individual equivalent widths are comparable when detected,
the overall absorption is discrepant.  To examine the overall
absorption, here we present equivalent widths per unit redshift
close to galaxies, including all LOS, both detections and non-detections.

\begin{figure*}
 \subfigure{\setlength{\epsfxsize}{0.9\textwidth}\epsfbox{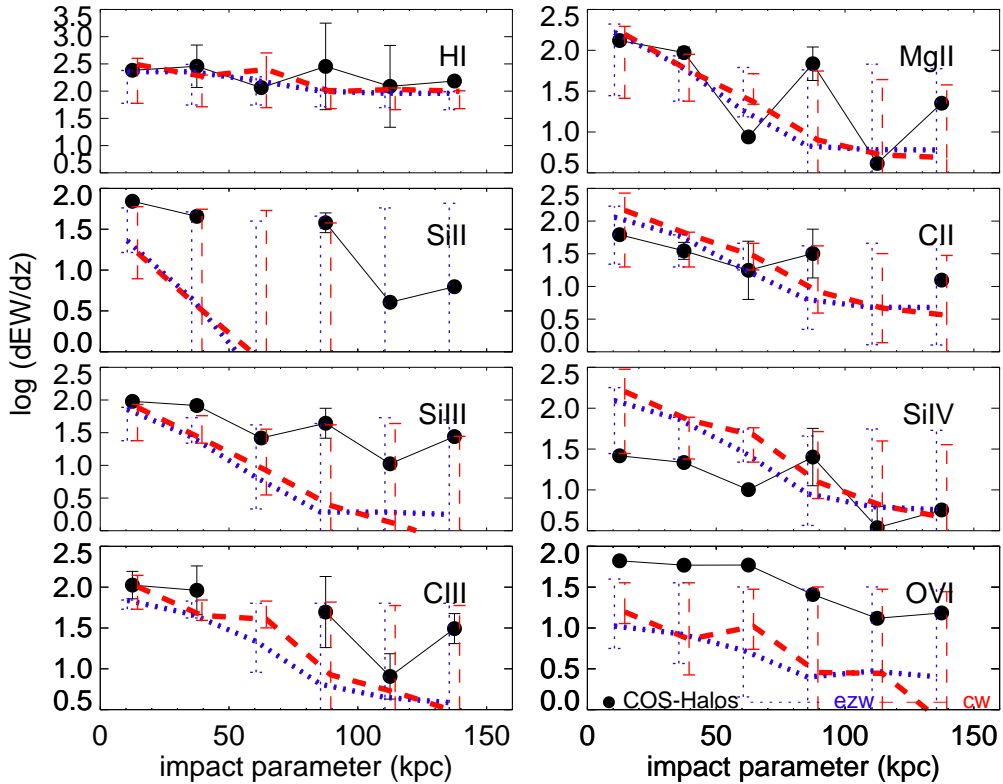}}
% \subfigure{\setlength{\epsfxsize}{0.8\textwidth}\epsfbox{N.jess.dec9b.ps}}
\caption{Summed equivalent width divided by total path length (dEW/dz) over all
LOS at a given impact parameter, as a function of impact parameter, binned in
25~kpc bins. Black points are from the COS-Halos survey, with error bars
averaged from well-constrained detections only, no limits or non-detections.
Blue dotted lines are for the ezw model and red dashed lines for the cw
model. Both models have range bars showing 16\%-84\% of all model LOS in a given impact parameter bin with detections above the thresholds given in Table \ref{table:thresh}.  
The errors on the model points have been horizontally offset slightly for
easier viewing. } 
\label{dedz}
\end{figure*}

Figure \ref{dedz} shows the equivalent width per unit redshift,
dEW/dz, versus impact parameter. We calculate dEW/dz by summing the
equivalent width (above the detection limits given in Table
\ref{table:thresh}) in all LOS probed in that impact parameter bin,
and dividing by the total redshift path length within $\pm 600$~km/s,
for all LOS with that impact parameter.  We bin both observed and
model points in 25~kpc bins. The observed COS-Halos points are shown
in black. For the purposes of this plot, observed points that are
upper or lower limits are assumed to be at their limit values; the
results do not change significantly if all upper limits were assumed
to be zero. The model values
are plotted as dotted blue lines for the ezw model and as dashed red lines
for the cw model.  The
bars on the model points are range bars, showing 16\%-84\% of the
model values for all LOS in that impact parameter bin (and is likely an
overestimate of the actual range at any given impact parameter in that bin).
The errors on the observed points represent the average
errors on all constrained detections in the 25~kpc bins.  These
errors do not include errors on LOS that are upper limits, lower
limits, or non-detections; hence the errors shown here are a lower
limit on the actual error. The errors on the model points have been
slightly offset horizontally for ease of viewing.

dEW/dz serves to diagnose whether our simulations are producing the
observed amount of \ion{H}{i} and metal ion absorption in the CGM
around these galaxies.  The \ion{H}{i} absorption is in excellent
agreement at all impact parameters.  For comparison, a random LOS
through the universe would typically yield dEW/dz$\approx$1.24, which is much
lower than the CGM typical value of several hundreds.  Hence the
agreement with \ion{H}{i} is non-trivial, and shows that there is
both the correct amount and the correct radial distribution of
neutral gas in the simulations.

Turning to the metals, \ion{Mg}{ii} and \ion{C}{ii} are well-reproduced
in the model, with the observed points being within the $16-84\%$
range of the simulation predictions. The models produce drastically too little
\ion{Si}{ii} and \ion{Si}{iii}; the models are off by as much as an order or magnitude. Conversely, the models produce too much \ion{Si}{iv} at low impact parameters, even outside the model ranges. The
simplest interpretation is that the models do not have as much cold,
dense gas in the CGM as observed, while having an over-abundance of warmer gas.  A physical interpretation might
be that our assumption of optically-thin gas fails to capture the
dense gas phase structure in the CGM.   Qualitatively, self-shielding
would be expected to increase \ion{Si}{ii} and reduce \ion{Si}{iv}.
However, we note that substantial self-shielding would greatly
increase the amount of \ion{H}{i} and \ion{C}{ii}, which at present
are in good agreement with the observations.  Clearly there is some disagreement
in the phase structure predicted in the models compared to that
observed, but the nature of this remains difficult to ascertain. Small features in the CGM which may give rise to absorption, such as interface regions and small, possible multi-phase substructures, may be below the resolution limit of the models presented here. 

For the mid-ion \ion{C}{iii}, the models produce slightly too little
absorption but generally the observed points are within the model
range values. For \ion{O}{vi}, however, the disagreement is much
more stark. While the difference in the {\it strength} of {\it
detected} lines (as shown in Figure \ref{ew}) between the models and
the observations is roughly a factor of 2, the difference in the {\it
total} absorption over {\it all} LOS is more like a factor of 5.
This difficulty in producing enough \ion{O}{vi} to match the
\cite{tum11} observations was also found in the {\sc Enzo} simulations
of \cite{hum13}.

The agreement between the models and the observations for
\ion{H}{i}, \ion{Mg}{ii}, \ion{C}{ii}, and to a lesser extent \ion{C}{iii}
is generally quite good given that
the models have not been tuned in any way to reproduce these data, and
suggests that the amount of CGM enrichment in both these simulations
is roughly in accord with observations. However, the mismatch in the silicon ions, as well as \ion{O}{vi}, is quite large. This is a significant failure, given both the degree of the mismatch, and the high quality of the 
observed \ion{Si}{iii} and \ion{O}{vi} data. 
The similarity in the
predictions of the two wind models is striking, particularly since
the amount of CGM metals and the phase space distribution of the
gas in these two models is rather different~\citep{for13b}, as we
will discuss later.

The trend in \ion{H}{i} is quite flat with impact parameter.  As
we discussed in the previous section, this partly owes to the
fact that a given impact parameter bin has a substantial range of
stellar masses.  Some of this likely owes to saturation, as \ion{H}{i}
at these equivalent widths are very much on the linear part of the 
curve of growth and hence fairly insensitive to variations in column density.
Nonetheless, the trend in \ion{H}{i} with $b$ is
even much flatter than in any of the low-ionisation metals.  At face value, this
suggests that the dropoff in metal ions at large radii owes either to gas becoming more ionised at higher impact parameter or to a dropoff in metallicity, rather than a dropoff in the
amount of neutral gas. 

In summary, the predicted total equivalent width per unit redshift
shows broad agreement with the observations, but with various interesting
discrepancies.  \ion{H}{i}, \ion{Mg}{ii}, \ion{C}{ii}, and \ion{C}{iii}
are in fair agreement with the observations, and \ion{Si}{ii}, \ion{Si}{iii},
and \ion{O}{vi} are significantly underproduced.  It is difficult
to identify a single physical variation that could even qualitatively
reconcile all these discrepancies while preserving the good agreements,
suggesting that further work is needed to discern their detailed
cause.  Radial dependencies are stronger for lower ions, except for
\ion{H}{i} that shows almost no decline with radius possibly owing
to saturation.

\subsection{Covering Fractions}

The covering fraction quantifies how patchy the CGM absorption is
in a particular ion.  Observations and simulations suggest that the CGM
has a complex, multi-phase structure, potentially with small clouds
embedded in warmer gas~\citep{tri08,kac12,for13b,cri15}.  The size and
fraction of material in these various phases remains poorly
constrained. \citet{chu03}, based on observed \ion{Mg}{ii} absorption around
galaxies, finds that this absorption generally arises in small clouds, i.e.
$2\le n_{H} \le 20$ cm$^{-3}$, line-of-sight sizes $1 \le D \le 25$~pc,
and masses between 10 and 1000 $M/M_\odot$.  In contrast, \citet{wer14},
from low-ionisation absorbers in COS-Halos, find a median cloud mass of ${10}^{7.6} M_\odot$.  Covering fractions thus provide
another test of models, and potentially a very stringent one if the
cloud sizes are well below what can be resolved in our (or most)
simulations \citep{cri15}. 

\begin{figure*}
 \subfigure{\setlength{\epsfxsize}{0.9\textwidth}\epsfbox{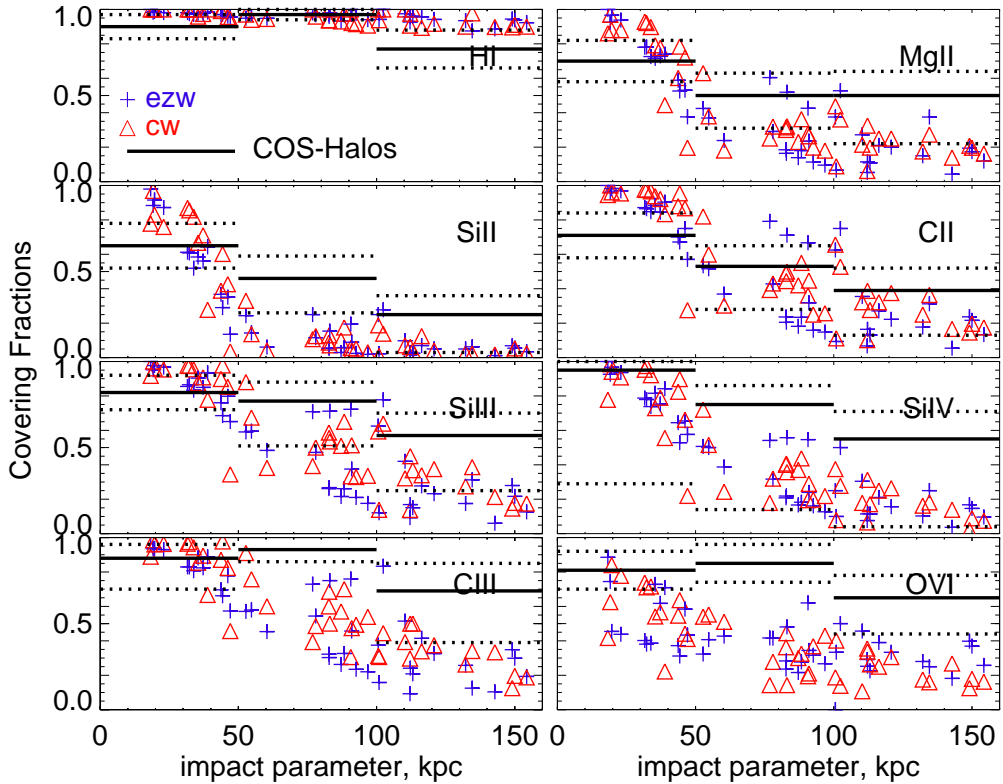}}
\caption{Covering fraction of COS-Halos survey data (black solid lines are
values, black dotted lines are errors), compared to values from the ezw
(blue crosses) and cw (red triangles) models. Observed values are in 50~kpc 
bins, and model points are not to show the downward trend with impact parameter.} 
\label{cf}
\end{figure*}

Figure \ref{cf} shows the covering fraction as a function of impact
parameter for our set of ions.  As in Figure \ref{ew}, we plot ions
from low to high ionisation potential. We define a covering fraction
as the fraction of LOS with a detection above a given ion-specific
EW threshold, corresponding approximately to the COS-Halos survey
detection limits, as listed in Table \ref{table:thresh}.  The
observed points are binned from 0-50~kpc,50-100~kpc, and 100-160~kpc.
The solid black line represents the median covering fraction, and
the dotted black lines represent the uncertainties (Wilson intervals). The models are
plotted as individual points (crosses for ezw, triangles for cw,
as before), because in the simulations we have multiple LOS at each
impact parameter and are able to define a unique covering fraction
at each impact parameter. In this manner one can get a sense of the
variability in covering fraction as a function of impact parameter,
albeit with the same degeneracy in stellar mass noted earlier. 

For \ion{H}{i}, the simulations almost always have unity covering
fraction above its detection threshold of 0.2\AA.  The covering
fraction in the COS-Halos survey is also very close to unity, but
formally slightly below.  This high \ion{H}{i} fraction was also emphasised in \cite{pro11}, and is independent of both galaxy
stellar mass and galactic star formation rate (not shown here), for
the range of these values covered by the COS-Halos survey. 
In fact
for all ions shown here, the covering fraction depends more strongly
on the impact parameter than either the star formation rate or
the stellar mass. The high covering fraction of \ion{H}{i} is noted
in the COS-Halos survey for both star-forming and passive
galaxies~\citep{tho12}.  We note that we are able to reproduce these
high \HI\ covering fractions in passive systems with our heuristic
quenching prescription (present in the ezw model but not the cw
model).

Further exploring Figure \ref{cf}, we see that for low ions
(\ion{Mg}{ii}, \ion{Si}{ii},\ion{C}{ii},and \ion{Si}{iii}) the models
generally are either at, or slightly above, the observations at impact
parameters of 0-50~kpc but drop below the observations at larger impact
parameters. For the mid ion \ion{Si}{iv}, the model values are below the
observed values in the 0-50~kpc bin, but within the errors. At
higher impact parameters the model covering fractions drop below
the observed errors.  For \ion{C}{iii}, there are model points below
the observed errors at all impact parameters. For the high ion \ion{O}{vi}
however, the model covering fractions are already too low at all
impact parameters, with the match worsening at higher impact
parameters.  We note that the covering fraction of the high ion \ion{O}{vi}
does not decline as steeply with impact parameter as the lower ionisation
species, either in the observations or in the models. We also emphasise the
similarity of the ezw and cw model points. While there is scatter
in the model points, there is not a single impact parameter bin in
a single ion where one model is consistently different from the other.
Rather, both models match (or mismatch) the observations in similar ways.

The mismatch between the models and the observations has three possible origins:
there could be not enough metals in the simulation, the metals could
be in other ionisation states not shown here, or the metals could
be in the wrong spatial distribution, which could happen if the
metals were not mixed properly.  We note that in both Figure
\ref{dedz} and \ref{cf} the radial gradients of \ion{Si}{II}, \ion{Si}{iii},
\ion{C}{iii}, and \ion{O}{vi} are steeper in the models than in the
observations. This suggests the models do not produce enough metals in the
outskirts of halos to match the COS-Halo observations. It
is also possible that some of the metals are not in the correct
ionisation states: note our overproduction of \ion{Si}{iv} at small
impact parameter but an underprediction of \ion{Si}{ii} and
\ion{Si}{iii}. We also slightly overpredict \ion{C}{ii} but
underpredict \ion{C}{iii} at low impact parameters, although the
discrepancies are generally within the error bars of the observations. It
is likely a combination of these factors that account for the mismatch,
as we discuss in further detail in \S\ref{sec:ionratios} and
\S\ref{sec:baryonfractions}.

In our simulations, we do not mix metals into CGM gas from the
particles that are ejected from the ISM, and hence it is perhaps
not surprising that there is too little metal mixing in our models.
It is not obvious how much mixing should occur; again, it is
challenging to model mixing of outflowing gas properly.  Nonetheless,
it is interesting that the covering fraction can potentially provide
a good constraint on this crucial physical process affecting how the outflows
and the CGM interact.

\subsection{Dependence on the Specific Star Formation Rate}
\label{sec:ssfr}

A key feature of COS-Halos is that it includes a significant sample
of passive galaxies, and hence can examine trends of CGM absorption
with star formation rate.  Current models broadly suggest that passive
galaxies are surrounded by hot gaseous halos~\citep[e.g.][]{gab14},
unlike star-forming galaxies that are located in ``cold mode"
halos~\citep[e.g.][]{ker05}.  Hence one might expect trends in the 
strength of various ions as a function of specific star formation
rate.

We examine the dependencies on sSFR by investigating equivalent
widths versus impact parameter, separated by sSFR. As noted earlier,
\cite{tum11} found that blue galaxies generally have higher equivalent
widths of \ion{O}{vi} than red galaxies, regardless of impact
parameter or galaxy mass.  In contrast, \citet{tho12} found that
the \HI\ properties in the CGM of red and blue galaxies are only
marginally different, and \citet{wer14} similarly found that
low-ionisation absorption is also not greatly dissimilar.  We
use our ezw model, which includes our heuristic quenching prescription,
to examine how well our simulation reproduces these observations.

\begin{figure*} %[h]
 \subfigure{\setlength{\epsfxsize}{0.9\textwidth}\epsfbox{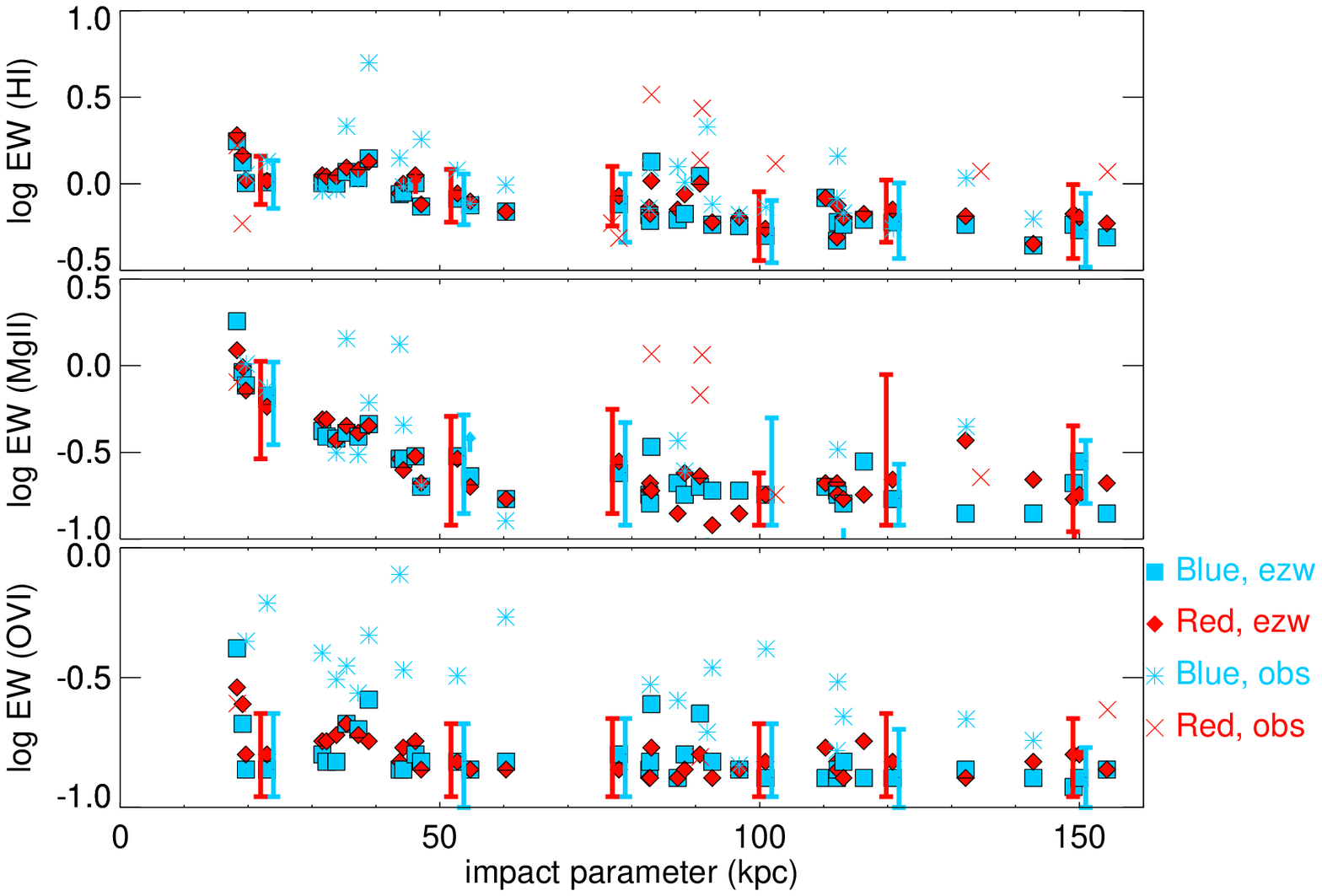}}
%\subfigure{\setlength{\epsfxsize}{0.6\textwidth}\epsfbox{cw.jess.nov25REDBLUE.ps}} 
\caption{Equivalent width vs. impact parameter of \ion{H}{i}, \ion{Mg}{ii}, and 
\ion{O}{vi} in the ezw model. Red model galaxies are shown as red diamonds,
blue model galaxies as blue squares, red observed galaxies as red x's, and blue
observed galaxies as blue asterisks. Red and blue are separated by 
a ${\rm sSFR}=10^{-11}~M_\odot/{\rm yr}$. Range bars show the 16-84\% values of the
model sightlines at fixed impact parameter, as in Figure \ref{ew}, but are larger here because the sample has been
divided.} 
\label{redblue}
\end{figure*}

In Figure \ref{redblue} we divide our model LOS -- which are initially
selected based only on impact parameter and stellar mass --  into LOS
around high- and low-sSFR galaxies (above and below $10^{-11}M_\odot/$yr).
To focus the discussion, we examine only \HI, our lowest metal ion
\ion{Mg}{ii}, and our highest ion \ion{O}{vi}; the trends for
intermediate ions are interpolatable from these.

We are able to reproduce the observed high \HI\ covering fractions
in passive systems~\citep{tho12} despite our heuristic quenching
prescription. In fact there is little difference in the equivalent
width of \ion{H}{i} in model red and blue galaxies (top panel),
which is consistent with what is seen in the observations (blue
stars and red crosses).  Hence our quenching model does not remove
\HI\ from halo gas, even though it adds substantial energy to the halo
by heating the ISM gas to temperatures that are well in excess of the virial
temperature.
We are not claiming that this lends physical validity to our simple
quenching model, but rather that such observations could provide interesting
constraints on more physically-motivated models of
quenching~\citep[e.g.][]{gab13}.  A similar trend is seen in
\ion{Mg}{ii} (middle panel), where our simulated red and blue
galaxies have similar absorption strengths.  In the observations,
if anything there seems to be more \ion{Mg}{ii} around red galaxies.
Our results suggest that whatever quenches the central galaxy in
more massive halos should have a minimal effect on the cool CGM gas
that gives rise to \HI\ and \ion{Mg}{ii}.

For \ion{O}{vi}, the situation is different. The observed strength
around blue galaxies are higher than around red galaxies \citep{tum11},
three of which
are upper limits off the bottom of the plot.  In
contrast, this difference is not present in the simulations: there
is no clear difference between the equivalent widths of \ion{O}{vi} in
red and blue galaxies.  Hence, the underprediction of \ion{O}{vi}
in the models seen earlier is specific to blue galaxies, while the
models greatly overpredict \ion{O}{vi} for passive galaxies.

This shows that, despite ejecting gas at high temperatures from the
ISM, our quenching model probably does not properly heat the
surrounding CGM gas to prevent cooling through the \ion{O}{vi}
temperature regime.  Our implemented quenching model
does not prevent accretion by keeping surrounding gas hot,
but only ejects the gas once it gets in the galaxy.  \citet{gab14}
argued that keeping the hot halo gas near the virial temperature,
well above where it would be in the \ion{O}{vi} phase, provides a
successful model to produce the quenched galaxy population as
observed.  Implementing such a model into our simulations would
likely have quite a different impact on \ion{O}{vi} absorption, as
it would explicitly prevent gas from cooling.  It remains to be
seen whether such a model would concurrently over-suppress \ion{H}{i}
and low ion metal absorption; we leave this investigation for future
work.  For now, we conclude that our implemented form of quenching
is inconsistent with CGM metal observations.  Clearly, matching both the
low ions and \ion{O}{vi} in the CGM of passive galaxies represents
an important constraint on the thermodynamics of quenching. Although passive galaxies in our simulations are not currently ejecting winds, we show in \S 4.3 that the CGM absorption in our
models is dominated by gas that was ejected well before the epoch
of observation, largely decoupling the CGM predictions from the
current star formation rate.

\subsection{Ion Ratios}
\label{sec:ionratios}

Ion ratios provide a complementary test for the physical conditions
of the absorbing gas.  Ratios between metal ions and hydrogen provide
an estimate of the metallicity in the gas traced by that ion, while
ratios between ions of the same element provide a constraint on the
ionisation conditions of the gas.  Given our current analysis, we
can only study ratios of ion equivalent widths summed over 600~km/s
intervals, which likely erases some of the detailed information
about physical conditions through the CGM; trying to subdivide these
into components introduces great sensitivity to the exact noise
level and fitting procedure, which is beyond the scope of this work
to consider.  Nonetheless, we will see that summed ion ratios still provide 
interesting constraints on CGM properties in our simulations.

\begin{figure*}
  \subfigure{\setlength{\epsfxsize}{0.9\textwidth}\epsfbox{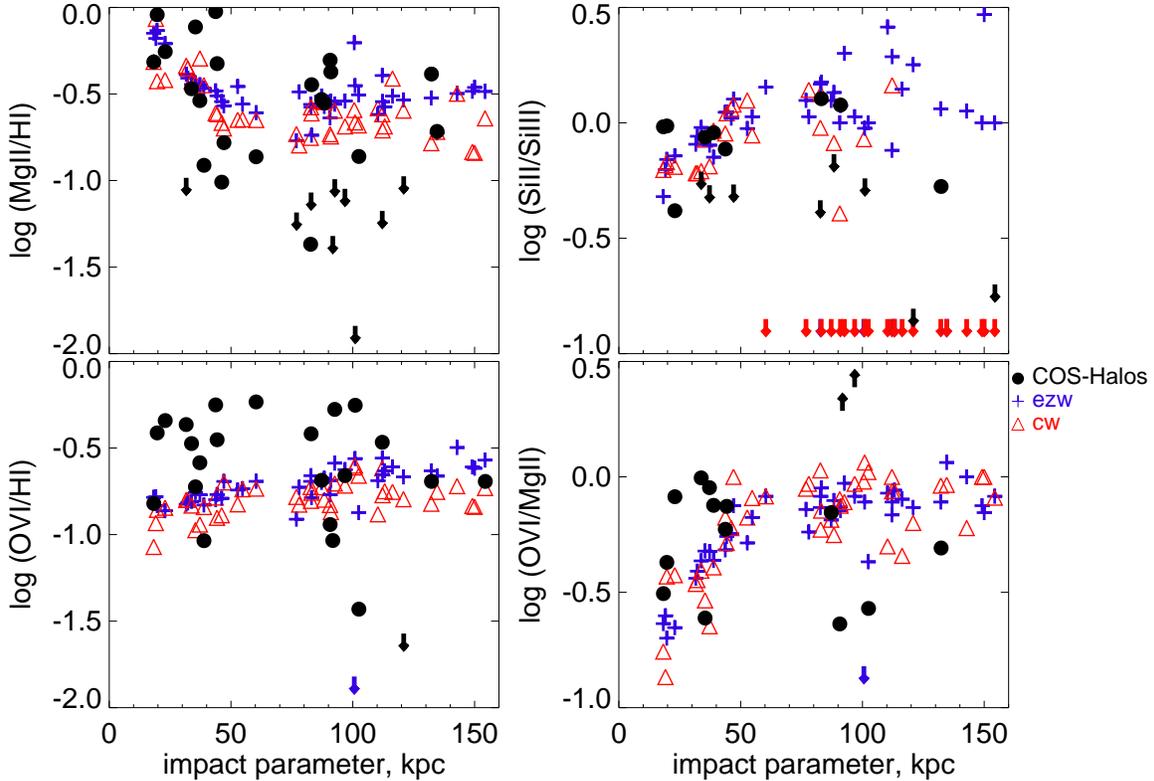}}
\caption{Comparison between the log of the ratio of equivalent width in the
COS-Halos survey (black symbols: circles for constrained values, downward
arrows for upper limits, upward arrows for lower limits), as well as for the
ezw model (blue crosses) and the cw model (red triangles). Coloured downward
triangles show where the model ratios are zero. Left panels show various metals 
vs. \ion{H}{i}, probing metallicity. Right panels show metal ratios, probing
physical conditions.} 
\label{ionratio}
\end{figure*}

Selected ion ratios are shown in Figure \ref{ionratio}. As before,
COS-Halos survey data is plotted in black, and model points (ezw:
blue crosses, cw: red triangles) represent the median value of
the ratio at each impact parameter. For the observed points, downward
facing arrows indicate that the ratio is an upper limit (meaning
an upper limit divided by an actual value, or an actual value divided
by a lower limit), upward facing arrows indicate the ratio is a
lower limit, and circles indicate an actual value. We do not plot
values that are poorly constrained, such as a limit divided by a
limit.  For the model points, coloured downward arrows show where
the ratio is zero, as little or none of the top ion is present.  In
the left panels, we plot the log of the ratio of equivalent widths of
a low and high ion relative to \ion{H}{i}, as a probe of the
metallicity of the gas traced by that range of ionisation potentials.
In the right panels, we show the ratio of two Si lines, as well as
a ratio of a low (\ion{Mg}{ii}) to high (\ion{O}{vi}) line.

The low ion ratio (\ion{Mg}{ii}/\ion{H}{i}) shows generally good
agreement with the observations, albeit with a scatter in the observations
that is much larger than in the models.  The scatter might arise owing to
local variations in ionisation conditions.  There is a slight hint
that the models produce slightly too high a ratio, particularly at
large impact parameter, indicating that the metallicity in the low
ionisation gas might be a bit too high. There is also a non-trivial
offset between the wind models, ezw is slightly higher than cw,
although the scatter in the observations is far too large to discriminate
between the models.  Still, this is one of the few aspects in
which any statistically significant difference can be seen between
the ezw and the cw models.

The predicted \ion{O}{vi}/\ion{H}{i} is considerably lower than
most of the observed points.  The discrepancy is comparable to that
seen in the individual EW measurements of \ion{O}{vi} vs. impact
parameter in Figure~\ref{ew}.  These discrepancies suggest that the hot diffuse phase in our models is
either insufficiently enriched with metals or has an incorrect temperature
structure that shifts oxygen to higher or lower ionisation states. These trends hold true for
both wind models (though ezw produces slightly higher ratios than
cw), indicating that it is not specific to which wind model we use,
but rather may be indicative of an overall failure in the way we
implement winds, namely by ejecting cool, unmixed ISM particles.
Ejecting hotter or over-enriched gas that mixes would qualitatively
be expected to move towards better agreement with the observations.
Nonetheless, we emphasise that the results are within a factor of $\approx$~3 of the observed ratios,
which is already a non-trivial success.  These ion ratio comparisons
further refine and highlight issues with our models, and demonstrate
the ability of the COS-Halos survey to provide important insights
into galactic outflow and ionisation processes in the CGM.

The right panels of Figure \ref{ionratio} plot different metal ions
against each other, mitigating the dependence on metallicity and
allowing us to test physical conditions of the gas more directly.
However, the observational data are more limited, as it is not common
to find absorbers with multiple ions detected.
The ratio of \ion{Si}{ii}/\ion{Si}{iii} at low impact parameters
shows an overall trend of increasing ratio with impact parameter,
and both models and are in good agreement with the observations from $\approx$
0-50~kpc. Beyond 100~kpc, the ezw model has a higher ratio than
the observations, indicating cooler gas than observed.  Meanwhile, the cw
model shows a very low \ion{Si}{ii}/\ion{Si}{iii} ratios, typically
having essentially no \ion{Si}{ii} and hence dropping off this
plot at these large impact parameters (shown as coloured downward
arrows).  The scatter is less than in the ratios versus \ion{H}{i},
suggesting that the ionisation conditions giving rise to particular
ions are more uniform in the CGM.

The \ion{Mg}{ii}/\ion{O}{vi} ratio shows a large scatter in the
observations, indicating that these two ions arise in different
phases that do not trace each other well.  There is, perhaps
surprisingly, no clear trend with impact parameter, as one might
expect for a simple model in which the low ion arises only near the
central galaxy and the high ion arises throughout the halo~\citep{for13a}.
This suggests that the \ion{Mg}{ii} is occurring throughout the halo,
perhaps in satellite galaxies or cold dense clouds interspersed
with warmer diffuse gas.  Meanwhile, the models show a rising trend
at small impact parameter that flattens at larger impact parameter.
The median ratio does not show a large scatter, suggesting that \ion{Mg}{ii} and \ion{O}{vi} are appearing together more often
in these models than in observations.

In summary, by examining ion ratios we can isolate trends with
metallicity and ionisation state in the CGM as traced by various
ions.  While there is broad agreement between the models and the observations
for \ion{H}{i} and some low ionisation metal species, there is also significant
disagreement for higher potential ions,
particularly \ion{Si}{iii} and \ion{O}{vi}. Clear
trends emerge that suggest that the metallicity within certain
phases may be too high (for warm diffuse gas traced by mid ions)
or too low (for hotter diffuse gas traced by high ions).  Ion ratios
generally show moderate agreement with the observations,
albeit with smaller scatter
and perhaps some discrepancies at large impact parameter ($\ga
100$~kpc) where the low ions disappear in the observations.  The differences
between our two wind models are small compared to the differences
between the models and the observations, which may owe to the similarly
kinetic and cool ejection of material.  Nonetheless, the differences between wind models seen here are not seen in
most of the other statistics we have considered so far.  Ion ratios
thus provide new and interesting challenges to models of CGM gas,
and obtaining a larger and higher-S/N sample could potentially
provide the best constraints on the physical conditions in the CGM.

\subsection{Kinematics}

With spectroscopic redshifts for all its galaxies, the COS-Halos
survey provides complementary information on the kinematics of
absorbing gas relative to the systemic velocity of the host galaxy.
In principle, kinematics could distinguish outflowing gas from
inflowing, but in practice this division is not so clean~\citep{for13b}.
Nonetheless, the kinematics provide an additional constraint on CGM
models.  In this analysis we will no longer sum absorption within
$\pm$600~km/s of the galaxy, but instead consider individual
components to elucidate the kinematics.  We note that this introduces
additional sensitivity to how our profile fitter decomposes
the absorption into components, but here we really focus only on
kinematics and not line strengths, hence such a sensitivity should
not greatly impact our results.

\begin{figure*}
 \subfigure{\setlength{\epsfxsize}{0.9\textwidth}\epsfbox{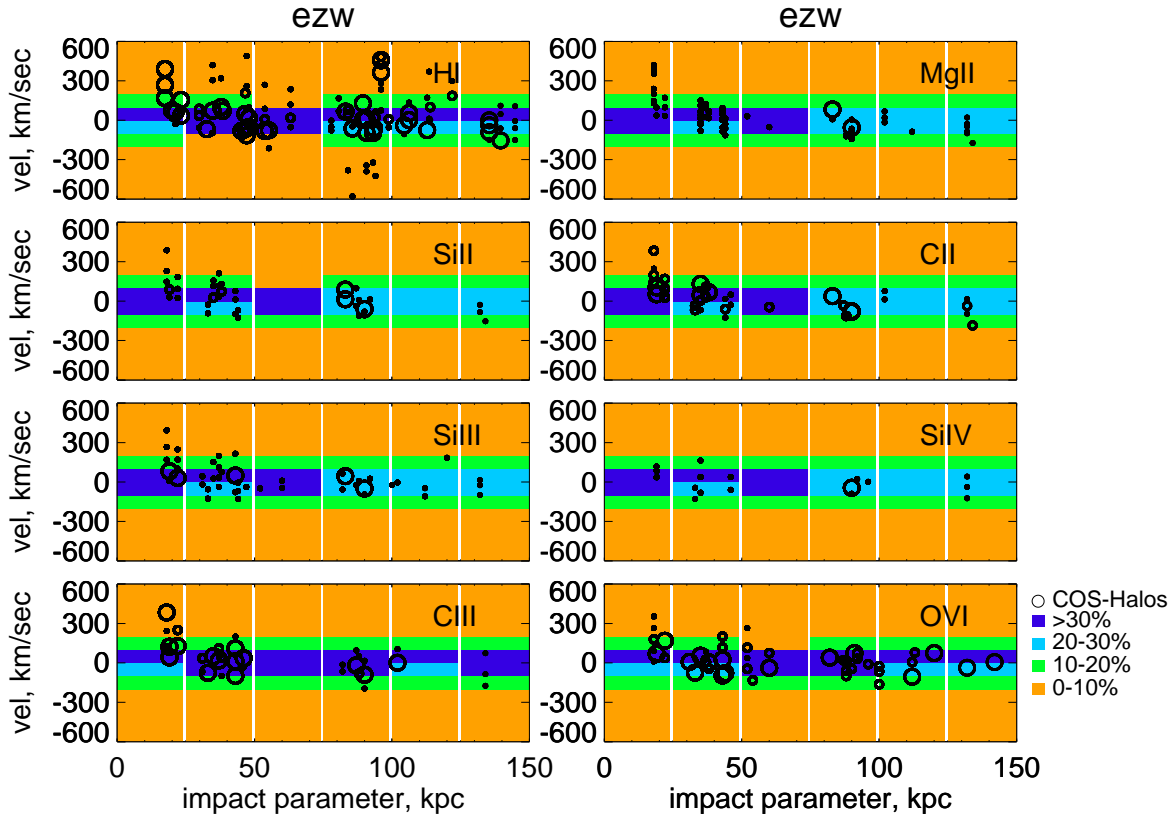}}
 % \subfigure{\setlength{\epsfxsize}{0.7\textwidth}\epsfbox{vel.cw.april11b.ps}}
 \caption{Distribution of column density along the line of sight. Ezw model data has been binned both in impact parameter and in velocity space (cw, not shown here, is similar). Dark purple blocks show that 30\% or more of the column density between -600 and +600 is concentrated in that velocity bin. Blue shows 20-30\%, green between 10-20\%, and orange less than 10\%. Overplotted as black circles are the COS-Halos observed points, scaled by the absorber strength. For the metals, large circles show ${\rm log}(N) > 14.5$, medium circles show ${\rm log}(N)=14.0-14.5$, and small circles show ${\rm log}(N) < 14.0$. For \ion{H}{i}, the scale is increased by 1: large circles show ${\rm log}(N) > 15.5$, medium circles show ${\rm log}(N)=15.0-15.5$, and small circles show ${\rm log}(N) < 15.0$.}
\label{velblock}
\end{figure*}

In Figure~\ref{velblock} we quantify from where in velocity space the
absorption originates in the observed data, and how this compares
to the model predictions.  We focus on the zw model; the cw model
gives very similar results.  To do so, we lay out a coloured
``chequerboard", where we bin our model data into 25~kpc spatial
bins and 100 $\kms$ velocity bins.  For each of those bins, we
calculate the total column density for all LOS with impact parameters
in that bin, for all absorption within $\pm$ 600 $\kms$.  We shade
each block by the fraction of the column density in that impact
parameter range that comes from a given velocity interval.  Dark
blue blocks indicate that more than 30\% of the total column between
$\pm$ 600 $\kms$ comes from that given velocity interval at that
impact parameter, light blue blocks indicate 20-30\%, green
10-20\%, and orange that less than 10\% comes from that velocity
interval. We note that these percentages are for that specific impact
parameter bin only, and not for the entire 0-150~kpc range, so each
vertical column of shaded blocks is independent.  We also plot the location of
each individual observed absorber (not summed over velocity, as elsewhere in
this work) as open circles whose sizes indicate the strength of the absorption.
For the metals, the large circles indicate ${\rm log} (N) > 14.5$, the medium circles
${\rm log}(N)=14.0-14.5$, and the small circles ${\rm log}(N) < 14.0$. For \ion{H}{i}, 
the scale is shifted up by one in the $\log$.

Generally, all the observed absorbers lie in the range where the
models predict a significant fraction of the absorption should lie. The observed
absorption also clusters strongly around the systemic velocities of the host
galaxies, which is
in line with the model predictions. The exception to this is that at
very low impact parameters there is an excess of absorbers with positive 
velocities, and for \ion{H}{i} and \ion{C}{iii} that absorption is stronger
than what is expected from the models. This owes to a single sightline that
happens to have strong absorption in multiple ions and lies at a positive
velocity with respect to its host galaxy.

Overall, both the simulations and the observations suggest that
most of the absorption occurs roughly within the virial velocity
of the galaxy's halo.  Only absorbers with the smallest impact
parameters show any sign of an outflow, and it is difficult to assess
its significance because it only occurs in one system (with several
absorbers within it), and there could be systematic uncertainties 
identifying the kinematic centre of any given system.  Otherwise,
the simulations broadly match the observed kinematics,
and do not indicate an abundance of strongly outflowing gas
at the present epoch, in general agreement with the outside-in
enrichment scenario of \citet{opp12}.

\section{Comparing the CGM in Wind Models}

In general, we have found rather modest differences in observational
tracers of absorption as probed by COS-Halos between our two wind
models.  Both models give similar predictions for the equivalent width of detected lines, total
equivalent width, and covering fraction of \ion{H}{i} and some of the
low ionisation metal lines, which broadly match data. Both models fail, most notably,
to reproduce the total equivalent widths of \ion{Si}{iii} and \ion{O}{vi},
and do so in similar ways. While we favour the ezw model for independent 
reasons, in the CGM observations presented here it typically does no
better or worse than the constant wind model in matching the observations.
This is somewhat curious, since earlier work \citep[e.g.][]{opp08,
for13a, for13b} showed significant observable differences between these
wind models in the physical conditions and enrichment of CGM gas. We
examine the reasons and implications of this surprising agreement here.

\subsection{Halo Mass-Stellar Mass Relationship}

A key difference relative to our previous works is that here we are
obtaining a galaxy sample that is matched in stellar mass, not halo
mass.  In our previous work we chose to match ezw and cw in halo mass since
we reasoned that the CGM is more tied to the halo, e.g. the chance of
metals escaping the CGM is tied to the full halo potential not just that of the
galaxy, and the total amount of CGM baryons should scale more with the
halo than with the galaxy.  However, COS-Halos has chosen galaxies based on
their stellar masses, and hence in this observational comparison we have
focused on matching this quantity.  As we argue here, it turns out that
the ``better than expected" agreement between the two models presented
here owes primarily to this choice, which has interesting implications.

\begin{figure*} %[h]
 \subfigure{\setlength{\epsfxsize}{0.7\textwidth}\epsfbox{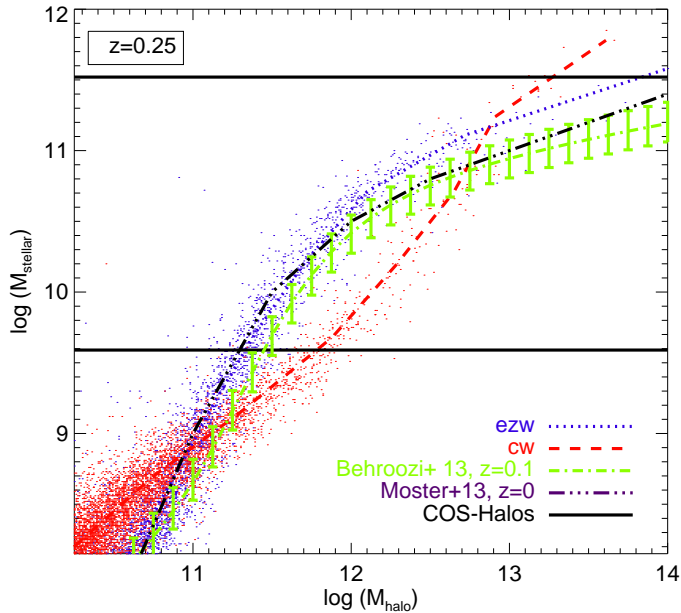}}
\caption{Stellar mass-halo mass relationship for central galaxies in the
ezw (blue dots, averaged as the dotted line) and cw (red dots, averaged as
the dashed line) at z=0.25. The stellar mass range of the COS-Halos data
set is bounded by solid black lines. Results from abundance matching are
shown: green dash-dot line ~\citep{beh13} and purple dash-dot-dot
line \citep{mos13}.} \label{hmsm} \end{figure*}

Figure \ref{hmsm} shows the stellar mass--halo mass (SMHM) relation
for our two wind models, with a running median for each (blue dotted
for ezw, red dashed for cw).  We include galaxies down to our stellar
mass resolution limit of $1.4\times 10^8 M_\odot$ (32 gas particle
masses). Solid black lines delineate the stellar mass range of COS-Halos.
The broken coloured lines show results from abundance matching analyses
for the closest available redshift: \cite{beh13} as the green dash-dot line,
and \cite{mos13} as the purple dash-dot-dot line.  The ezw model is clearly
a better match.  This simply reflects the fact that ezw (or its cousin,
vzw that does not have the energy-driven scalings at low masses or
quenching prescription at high masses) provides a better match to the
galaxy stellar mass function than cw~\citep{opp10,dav11a}.

The stellar mass range relevant to this work is ${\rm 10}^{9.5-11.5}
\msolar$ (see Figure~\ref{hmsm}), which from \citet{beh13} corresponds
to a halo mass range of ${\rm 10}^{11.4-15+}\msolar$. In much of this
range, the cw model's SMHM relation is lower than ezw's.  This mainly occurs
because the cw model ejects material at higher velocities in this mass range,
which reduces the amount of wind recycling that dominates the inflow at
these masses~\citep{opp10}.

In the present work, we compare the wind models at fixed stellar mass.
Hence for a given $M_*$, the cw model will predict a higher $M_{\rm
halo}$.  We can use this to qualitatively reconcile why \citet{for13a}
found large differences between the two wind models.  For example,
Figure 9 of \cite{for13a} shows that for a momentum-driven wind model
(similar to the ezw model used here), higher mass halos have {\it higher}
column density and hence EWs at a fixed impact parameter.  Figure 14
of that same work shows that for a fixed halo mass, the constant wind
model has a {\it lower} column density at fixed impact parameter. Therefore, at
fixed stellar mass, given that the constant wind model has higher halo
masses than ezw, this would result in a higher column density at fixed
impact parameter.  The differences in absorption strength owing to the
wind model are thus offset by variations owing to halo mass.  As a result,
for the observables presented above, the two wind models match much better
at fixed stellar mass than at fixed halo mass.  This then propagates into
all the various other statistics, such as covering fractions and dEW/dz.

The interesting physical implication of this is that our models predict
the CGM metal absorption properties are more tied to the stellar mass
than to the halo mass.  On the one hand, this is surprising since the CGM
represents halo gas, which one might think would be more closely tied
to the halo mass. That being said, the metals in the CGM are formed by the stars
in the galaxies and then ejected, so the closer relationship is perhaps
not unexpected.  In \citet{fin08} and \citet{dav11b} it was argued that
the metals retained within the ISM, as traced by the mass-metallicity
relation, are strongly governed by outflows, whose properties must be
tied most closely to stellar mass in order to give rise to a tight
stellar mass-metallicity relation~\citep[e.g.][]{fin08,dav11b,pee11}.
Evidently, in these simulations, the metals deposited into the CGM are
likewise strongly governed by outflows that are more closely tied to
the stellar mass.

We emphasise that this similarity in predictions for ezw and cw does
{\it not} imply that current CGM observations are unable to rule
out any outflow models.  While we have varied the free parameters
in our particular implementation of outflows, there are many other
implementations of outflows in simulations that may result in quite
different CGM properties.  These include models that over pressurise the
ISM by local supernova heating~\citep{hum13,she13,ros14}, drive winds
through a combination of radiation pressure and supernovae~\citep{hop13},
and include early stellar feedback driven by stellar winds and
radiation~\citep{sti13}.  Compared to these, our simulations make some
rather extreme assumptions, namely that the ejected particles are not
initially heated (though they can be shock heated by interactions with
the ambient halo gas), and that our metals do not mix with the ambient gas.
It appears that these assumptions may be more critical to setting the CGM
enrichment and phase structure than the details of the assumed wind speed
and mass loading factor.  For instance, \citet{hum13} found that their
AMR-based models did not fare well against CGM observations, except with
rather extreme assumptions about outflow energetics.  We look forward
to detailed comparisons of other simulations, similar to what we have done here, in order to better assess how
the physics driving outflows manifests in CGM absorption line data.

\subsection{Baryonic Fractions within Halos} \label{sec:baryonfractions}

\begin{figure*} %[h]
 \subfigure{\setlength{\epsfxsize}{0.7\textwidth}\epsfbox{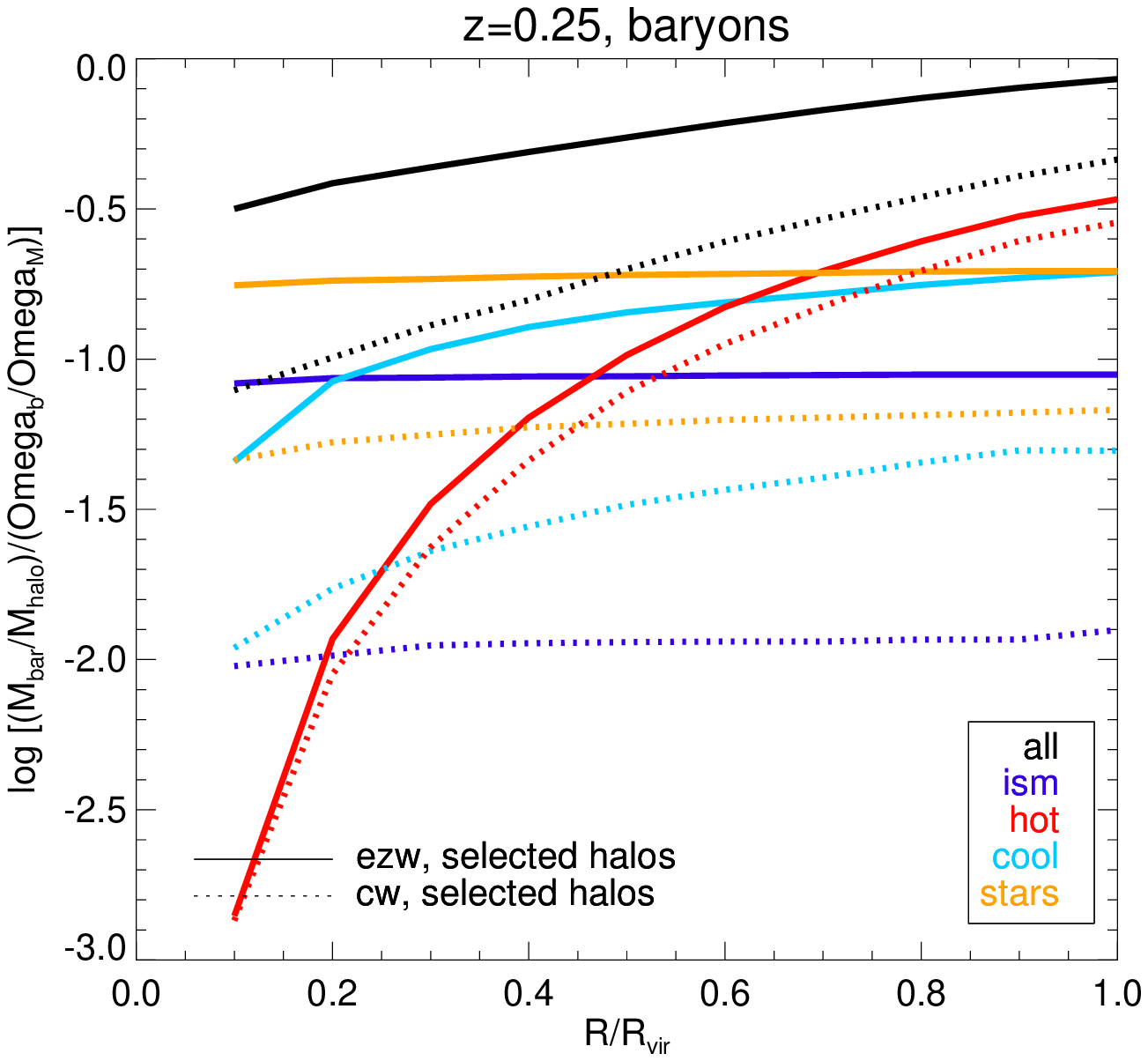}}
  \subfigure{\setlength{\epsfxsize}{0.7\textwidth}\epsfbox{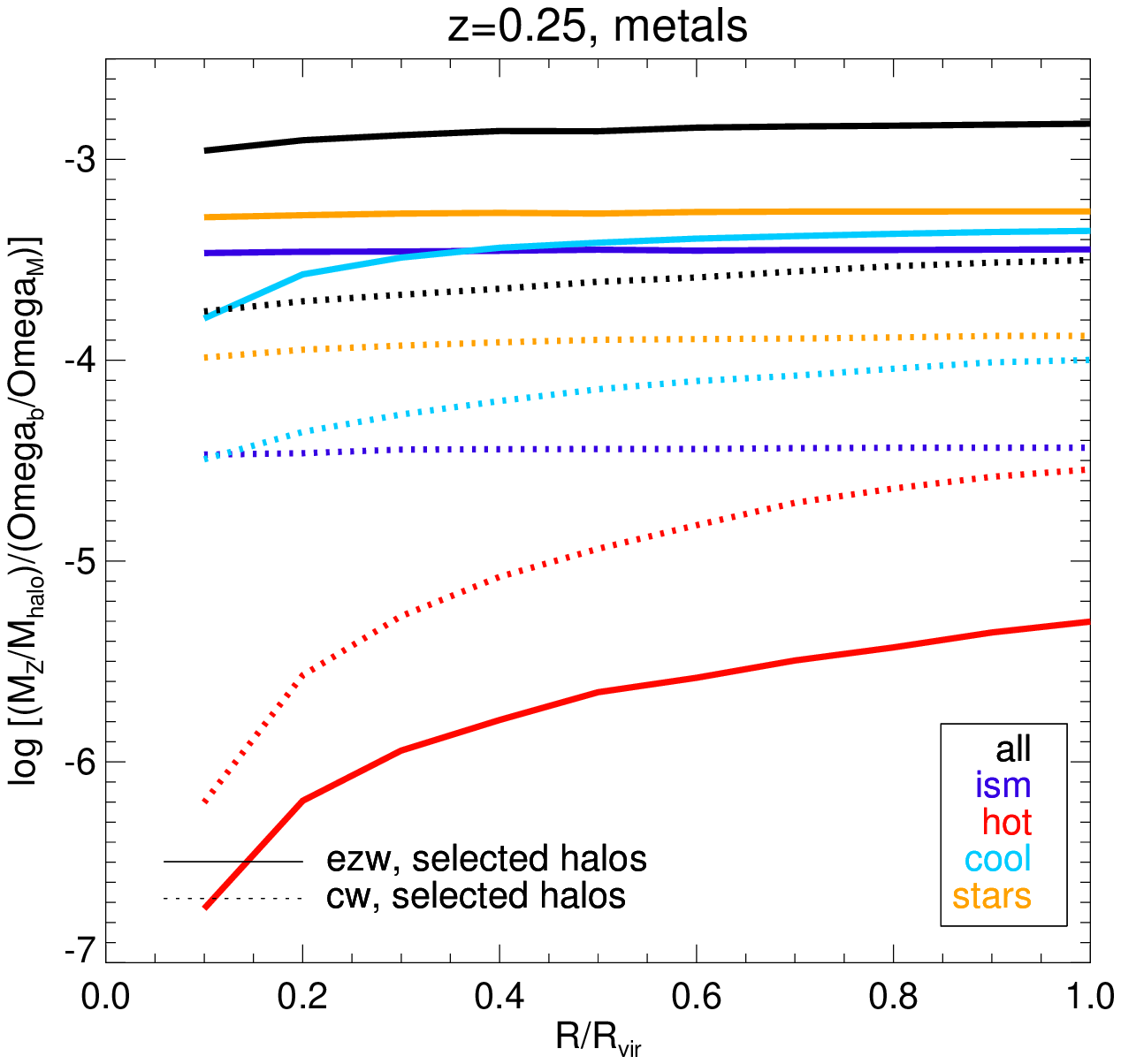}}
\caption{Baryonic content (top panel) and metal content (lower panel)
of our COS-Halos matched galaxies for ezw (solid) and cw (dotted) wind models
as a function of enclosed radius. The black lines show the total baryon fraction
for the ezw and cw models. Dark blue, red, light blue,
and orange lines show the fractions of ISM (star-forming) gas, hot gas
($>$ $10^{5}$~K), cool gas ($<$ $10^{5}$~K), and stars, respectively.}
\label{halofrac} \end{figure*}

The two wind models considered here predict similar CGM absorption line
properties, despite differences in the amount and physical state of CGM
gas. We examine those differences here.  

Figure \ref{halofrac} shows the {\it cumulative} enclosed baryonic fraction
(top panel) and metal fraction (lower panel) within halos as a
function of radius, for our COS-Halos matched sample of galaxies.
The upper panel shows the total baryonic mass fraction divided by
the cosmic baryon fraction (i.e. $\Omega_b/\Omega_m$).  Unity
indicates a cosmic mean baryon fraction, and shows approximately
what models would predict in the absence of outflows~\citep[in
detail, it would be slightly higher than unity at these masses; see
e.g.][]{dav09}.  The solid black line shows the total fraction of
baryons (stars+gas) in halos for the ezw model, as a function of
enclosed radius from the galaxy. Even at the virial radius, this
is less than unity, showing the efficiency of outflows in driving
out baryons. The dotted black line shows the baryon fraction for
the cw model, which is substantially lower than that of the ezw
model. Owing to higher wind speeds in this halo mass range, the
constant wind model carries more of its baryons out of halos into
the IGM than ezw.  Both models push baryons out preferentially from
the central regions, where most of the star formation (and hence
outflow generation) occurs. This leaves a large component of gas-phase
baryons in the CGM. 

For the ezw model, CGM gas (hot+cool phases) makes up 65\% of all
baryons inside the halo. This is broadly consistent with estimates
of baryonic mass derived from observations \citep{wer14}. The cw
model predicts many fewer baryons, particularly in the cool phase.
However, it is important to note that in converting observable
quantities into estimates of mass, assumptions must be made about
the physical conditions in order to obtain ionisation corrections.
The consistency of ezw here, then, may simply reflect the similarity
in physical conditions in the ezw model to those assumed in
\citep{wer14}. An assumption of a higher proportion of hot gas, as
present in the cw model, would alter the conversion of observable
quantities to mass.  Hence this should not be regarded as a success
of ezw relative to cw, but may provide guidance on what assumptions
should be made in order to obtain CGM mass estimates from absorption 
line data.

The ISM of galaxies in the cw model contains an order of magnitude less baryons
than in the ezw model,
and about one-third as many stars.  The reasons for this are two-fold:
more gas is driven out of the ISM, making less available for star
formation and strong outflows can actually heat the surrounding region thereby
preventing the accretion of fresh gas that would otherwise replenish the
loss~\citep{opp10,van11}.  The radial profile of the ISM and stellar
fractions are both quite flat, since these phases are concentrated in
the galaxy's central regions.

More directly relevant to CGM absorption lines is the cool CGM baryon
fraction, which is about three times higher in the ezw model than in the cw
model and shows a modest increase with radius.  In contrast, the hot ($T>10^5$K)
baryon fraction is not very different in the two models, and increases
strongly with radius.  The fact that the \HI\ equivalent widths
are much less than a factor of three different in these two models (see
Figure~\ref{ew}) is, therefore, curious and suggests that the global CGM
cool gas content is not straightforwardly traced by \HI\ absorption.  However, 
this again may owe largely to saturation effects in \HI.

The lower panel of Figure \ref{halofrac} shows the metal fraction in
various halo gas phases. Here we have plotted the metal mass fraction
enclosed within a radius $R/R_{\rm vir}$, divided by the cosmic baryon
fraction (i.e. $\Omega_b/\Omega_m$) for ease of comparison with the upper
panel. The constant wind model has a lower enclosed metal fraction, since
more of the metals have been driven out of the halo by outflows and since
it made less stars. Of the
metals that are present in the cw model, more are hot. For a sense of 
the metal content derived from observations of COS-Halos galaxies, we refer the reader to
\cite{pee14}, but again we note that conversions between observables and metal mass 
require an assumption of physical conditions.  Hence it is not straightforward
to use such inferred measures to discriminate between wind models.

In summary, we see that our two wind models distribute baryons and metals
differently, even though they generally give similar observational
absorption line properties.  This suggests that the total amount of CGM gas and metals
is not sufficiently constrained by the current observations to distinguish 
between competing outflow models.

\subsection{The dynamical state}

\begin{figure*} %
%\subfigure{\setlength{\epsfxsize}{0.9\textwidth}\epsfbox{EW.catjess.feb2.ps}}
\subfigure{\setlength{\epsfxsize}{1.05\textwidth}\epsfbox{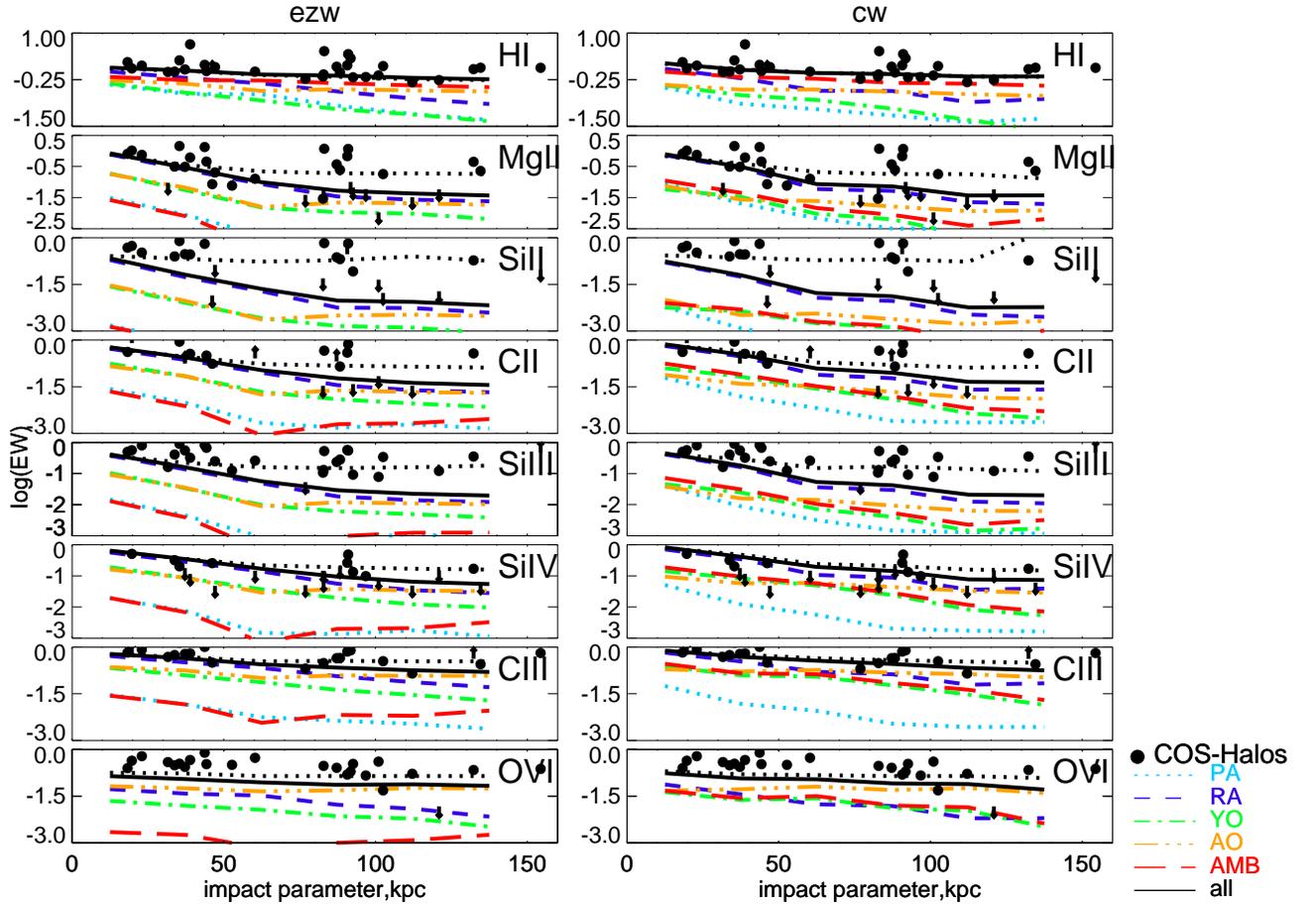}}
\caption{Equivalent width (\AA) versus impact parameter for the COS-Halos data
set (black symbols both panels), as well as for the ezw (left) and cw (right)
models, for various ions ordered from low to high ionisation energy. The
model points represent the mean value of all LOS in 25~kpc impact
parameter bins. Broken coloured lines correspond to pristine accretion
(PA; dotted light blue), recycled accretion (RA; dashed purple), young outflows
(YO; dash-dot green), ancient outflows (AO; dash-dot-dot orange) and ambient
(AMB; long dash red). Solid black lines include all particles that lie along
the LOS, regardless of category, and in contrast to Figure \ref{ew} they include LOS below the detection threshold.} 
\label{EWcat} 
\end{figure*}

A key goal of CGM absorption line work is to be able to distinguish
whether a given absorber arises from inflowing, outflowing, or ambient
material~\citep[e.g.][]{bur13}. In \citet{for13b}, we studied general
observational diagnostics of the dynamical state of the gas in the
ezw model and found, for example, that low metal ions tended to trace
recycled accretion while high metal ions traced ancient outflows (both defined below).  Here, we
expand on this past study by examining and contrasting the ezw and cw wind
models, while focusing on our simulated LOS that mimic the COS-Halos data set.
In this way we aim to develop intuition about how COS-Halos and similar
CGM absorption line data sets are able to trace the baryon cycle.

The first step is to characterise inflowing and outflowing gas in the
simulations, for which we follow \citet{for13b}. Briefly, we separate
CGM gas particles into five different kinematic categories, utilising
the past and future information we have within the simulations.  Since we
are interested in CGM and not ISM, each of these categories excludes
gas that is in the ISM at $z=0.25$.  The categories are:

\begin{enumerate}
\item {\bf Pristine Accretion}. This is gas that is accreting, meaning
it is not in the ISM of a galaxy at $z=0.25$ but will either end up a the
galaxy (as an ISM particle or star particle) or pass through
a galaxy and be ejected as a wind particle by $z=0$. Pristine accretion
is accreting gas that has not previously been in a wind.
\item {\bf Recycled Accretion}. This is gas that is accreting, as defined above,
but has been in a wind prior to $z=0.25$.
\item {\bf Young Outflows}. This is gas that was ejected in a wind ``recently",
relative to $z=0.25$. We define recently as 1 Gyr prior to $z=0.25$
($z=0.36$ for our cosmology), as this is roughly the
time a particle would take to leave the halo if it were simply launched 
in a wind, and was not slowed down by hydrodynamic interactions.
\item {\bf Ancient Outflows}. This is gas that was
ejected in a wind longer than 1~Gyr ago (before $z=0.36$).
\item {\bf Ambient}. This is gas that will not accrete onto a galaxy by
$z=0$, and has never been in a wind by $z=0.25$.
One can think of ambient material as
gas that is neither inflowing nor outflowing, and hence not participating
in the baryon cycle.  
\end{enumerate}

In Figure \ref{EWcat} we plot the EW versus impact parameter, similar to
Figure~\ref{ew}, except now we decompose the simulated equivalent
widths into contributions from the five dynamical categories above.
We calculate these in the manner described in \citet{for13b}, by making
new simulation outputs containing {\it only} those gas particles that
fit the definition of the various categories, rerunning LOS through them,
generating spectra, and fitting column densities.  The ions are ordered
from low to high ionisation energies, as in earlier figures. We also plot as
solid black lines results obtained including particles
from all categories and the ISM. The contributions from all the categories
(coloured lines) do not necessarily exactly add to the total (black
line), as they have been fitted separately.  We also show the COS-Halos
data for reference, as in Figure~\ref{ew}, including detections and
limits. To avoid clutter we do not include the observed error bars,
which are the same as in Figure~\ref{ew}.  Finally, we bin both the model
and the observed points in 25~kpc bins. 

Each of these new simulation snapshots is by definition sparser than the complete simulation, so it is no longer appropriate to use the medians of LOS with detections above a certain limit, as in
Figure \ref{ew}. To account for the sparser simulation boxes, in Figure \ref{EWcat} we now
plot the mean of all the model LOS at a given impact parameter. As it is
not appropriate to compare the mean of all LOS (as plotted here) to only
detections (as in the observations), we caution against comparing the
solid black line to the observed black points. The observed points are
shown here to simply guide the eye and provide a rough comparison. For
a more detailed, fair comparison, see Figures \ref{ew}, \ref{dedz},
and \ref{cf}. We will now examine the behaviour of these ions in the models, with the understanding that these models can have significant discrepancies with the observed values, particularly for silicon and \ion{O}{vi}. 

The top panels show \HI, in ezw (left) and cw (right).  The general
trend is the same as that identified in \citet{for13b}: at low impact
parameters, \HI\ predominantly arises from recycled accretion,
tracing dense gas that is about to re-accrete within several Gyr, while at
high impact parameters it arises mostly in pristine ambient gas.  At all impact
parameters there is a non-trivial contribution from ancient outflows,
which follows the ambient curve at a level that is set predominantly
by the ratio of the masses in the ambient and ancient outflow categories.
These trends are qualitatively true in both wind models.

There are, however, non-trivial differences between these wind models.
Relative to the ezw model, in the cw model the recycled accretion only dominates
at the very smallest impact parameters.  Also at smaller impact parameters in
the cw model, the contribution from ambient
gas is 2-3 times more than than that from ancient outflows, whereas in the ezw
model it is comparable.  Despite these variations, the total (black) line
is very similar in both cases, showing that the \HI\ equivalent width is
not able to, on its own, discriminate the dynamical state of the CGM gas.

Our lowest ionisation ion is \ion{Mg}{ii} and, again in accord with
\citet{for13b}, we find that recycled accretion is the dominant
contributor at all impact parameters.  Beyond 100~kpc, there is
an increasing fractional contribution from ancient outflows, but
this is still sub-dominant.  This is likely because \ion{Mg}{ii} absorption
(and low ions generally) drops off very quickly with impact parameter
\citep{for13a}; at large impact parameters the absorption is likely
coming from satellites, not the central galaxy. This trend is true in
both wind models, though cw at small impact parameters shows a small
contribution from ambient gas and young outflows.

The trends for \ion{Mg}{ii} are mimicked by the mid-ion \ion{Si}{iv};
this ion is still predominantly tracing halo gas that will accrete onto
the galaxy.  In contrast, \ion{O}{vi} shows a dominant contribution from
ancient outflows at all but the smallest impact parameters.  As argued
in \citet{for13b}, this is consistent with the idea that \ion{O}{vi}
comes from a diffuse halo around the galaxy, enriched over long periods
of star formation \citep{tum11,opp12}.  These trends are also fairly
similar in the cw versus the ezw models, which indicates that these
interpretations are
robust at least to the range of wind models that we probe here. Note that, as discussed earlier, ours are not representative
of all possible outflow mechanisms in current CGM simulations.

In summary, the most clear trend is that at small impact parameters,
recycled accretion dominates the absorption of low ions (including
\ion{H}{i}), and that ancient outflows dominates \ion{O}{vi} absorption.
Most \ion{H}{i}, meanwhile, arises from ambient gas at larger impact
parameters.  These trends mimic the expectations from \citet{for13b}.
Hence the broad intuition developed in our previous work can be used
to roughly interpret absorption line observations from COS-Halos and similar
studies, even though they may be less clear owing to a conflation
of impact parameter and halo mass trends.  Our two wind models show
qualitatively similar behaviour, indicating that at least within the
class of outflow models considered here these results are robust. 
The dominance of recycled accretion for
low ions and ancient outflows for \ion{O}{vi} explains why our model predictions
for star-forming and passive galaxies can be so similar: the absorbing
metals are predominantly ones that were ejected long ago, and they are
not coupled to the current star formation rate of the central galaxy.
Thus, a quenching scheme that shuts off star formation but does not
directly influence gas already in the halo has relatively little impact
on CGM observables. Similarly, since the ambient gas dominates the \HI\ 
absorption, one expects little difference between star forming and passive
galaxies, except perhaps at small impact parameters where recycled accretion
becomes important.

\section{Summary}

We have compared our cosmological hydrodynamic simulations including two
galactic outflow prescriptions with \HI\ and metal absorption line
observations in the CGM from the COS-Halos survey, to (i) assess how
well current models that self-consistently enrich the CGM and IGM fare
against state of the art CGM observations; (ii) determine if available
observations are able to distinguish between outflow models; and (iii) provide
interpretations about the physical and dynamical state of CGM gas as
observed by COS-Halos.  To accomplish this, we extracted LOS around
simulated galaxies that match the COS-Halos sample closely in stellar
mass and impact parameter, analysed them in a manner consistent with
the observations, and made comparisons to direct observables, namely covering fractions and 
equivalent widths as a function of impact parameter and specific star
formation rate and the kinematics of the absorbers relative to the
host galaxy.  In particular, we compared to COS-Halos data tracing
\ion{H}{i}, \ion{Mg}{ii}, \ion{Si}{ii}, \ion{C}{ii}, \ion{Si}{iii},
\ion{Si}{iv}, \ion{C}{iii}, and \ion{O}{vi} in 44 sightlines out to an
impact parameter of 160~kpc around galaxies with stellar masses between
$10^{9.5}\la M_*\la 10^{11.5} M_\odot$ at $z=0.25$.

Our main conclusions are as follows:

\begin{enumerate}

\item Our favoured wind model, ezw, is in broad agreement with key
absorption line observables from COS-Halos for \ion{H}{i} and some
low ionisation metal lines. In particular, for these ions we find good
agreement in the observed equivalent widths versus impact parameter, the
total equivalent width per unit path length, and ion covering fractions
above the COS-Halos detection limit. This model is also in broad agreement with
ion ratios and absorber kinematics relative to the host galaxy.

\item However, there are numerous discrepancies that hint at missing or
poorly represented physics in the simulations.  These include:

\begin{itemize}

\item Both wind models underpredict the two most commonly observed ions
from COS-Halos, namely the mid-ion \ion{Si}{iii} and high-ion \ion{O}{vi}.
For \ion{O}{vi}, the mismatch occurs at all impact parameters, while
for \ion{Si}{iii} the match is increasingly poor towards larger impact
parameters.  This mismatch occurs in both equivalent width per unit
redshift and the covering fraction.  In general, all the ions but
\ion{O}{vi} have a steeper gradient of absorption versus impact parameter
in the models than what is observed, likely suggesting that there are not enough
metals in the outskirts of halos in the models. We also note that both wind model simulations have the same resolution, and hence fail equally to reproduce any CGM structure below the resolution limit. 

\item The total amount of absorption in \ion{Si}{iv}, a mid ion,
is generally too high compared to observations, while the amount in high ion
\ion{O}{vi} is too low.  The ion ratios \ion{Si}{iv}/\ion{H}{i} and
\ion{O}{vi}/\ion{H}{i} echo these trends.  This is likely to be an issue
related to how much metals are deposited into different 
phases of the CGM, in the sense that our model puts too many metals in
$\sim 10^4$K gas and not enough in $\sim 10^5$K gas.
This occurs in both wind models, hinting that the underlying cause may
be the way in which we eject winds as cold, unmixing gas from the ISM.

\item The large scatter in the equivalent widths at a given impact
parameter for low ionisation lines is not well reproduced in the
simulations.  This suggests that local effects perhaps owing to local
ionisation sources and self-shielding might be necessary to fully explain
the distribution of these low ions.

\item While observations show a substantial difference in the
incidence of OVI absorption around star-forming and passive galaxies,
our simulations predict similar levels of absorption.  This discrepancy
may arise because our phenomenological quenching scheme suppresses star
formation in massive galaxies but does not directly affect halo gas.
Conversely, our models successfully explain the observed similarity of
low-ion absorption around star-forming and passive galaxies.

\end{itemize}

\item The constant wind model produces CGM absorption in all ions that
is surprisingly similar to that from the ezw model, despite substantially
different input wind scalings, which results in higher halo gas temperatures
and fewer cool baryons and metals.  It is not possible to distinguish
the cw model from the ezw model using the available equivalent width,
covering fraction, or kinematic data, although ion ratios do reveal some
differences. The lack of discrimination between the ezw and the cw models
is also surprising
given our previous results in \citet{for13a,for13b} that showed larger
differences between similar models.  We show that this owes largely
to the fact that previously we compared models at a fixed halo mass,
whereas here we compare at a fixed {\it stellar} mass.  The
CGM metal absorption properties in these models are thus more closely
correlated with stellar mass than halo mass.

\item Both the ezw and cw models show large contribution from recycled
accretion at low impact parameters ($\la 50$~kpc) in low ions including
\ion{H}{i}, and from ancient outflows in \ion{O}{vi} at most impact
parameters, in agreement with our previous results~\citep{for13b}.
These trends are robust for our simulations, suggesting that they
can be used to roughly infer the dynamical state of CGM absorbing
gas at least within the context of the probed outflow models.

\end{enumerate}

The agreement in \ion{H}{i}, \ion{Mg}{ii}, and \ion{C}{ii}  between
hydrodynamic simulations including galactic outflows and the most
recent observations of CGM gas offers encouraging support for the
baryon cycling paradigm. The disagreement between both wind models
and observations in \ion{Si}{iii} and \ion{O}{vi} suggests further work
is necessary in the simulations to get outflows to place metals in the
proper location and phase in the CGM.

While the COS-Halos observations of equivalent widths and covering
fractions cannot discriminate between the two wind models we consider
here, it is important to recognise that these span a relatively narrow
range of possible outflow models.  The ezw and cw models have different
mass loading factors and ejection velocities, but in both cases our
implementation ejects gas at the ISM temperature (usually $\sim 10^4\,{\rm
K}$) and does not mix metals from SPH wind particles into the surrounding
CGM. These assumptions are a limiting case, since some mixing must
certainly occur, and if supernovae are involved in driving outflows it is
likely that the ejected gas will be at least somewhat heated.  Other groups
have utilised fundamentally different algorithms for ejecting outflows,
for example by super-heating the ISM gas, which may give much larger
variations in CGM properties than either of the models we consider here.
We plan to consider a wider range of outflow driving mechanisms in the
future, so this work should be regarded as an initial investigation
that we hope will set the stage for a more comprehensive comparison
amongst all the groups working on models of the CGM and the baryon cycle.
The discrepancies between our models and current observations already
suggest that we must improve our models of outflow ejection, propagation,
and mixing.

Within the context of the models we consider here, our interpretations
of the COS-Halos observations are relatively insensitive to the details of how
we implement winds. For instance, the conclusion that low ion absorption
is dominated by enriched gas that will shortly fall into the galaxy,
while high ions arise in gas that was enriched at earlier times and is
stably located in hot gaseous halos, is quite generic for our models.
Of course, these interpretations may be different if we fundamentally
changed our wind driving mechanisms, as discussed above.  Nonetheless,
the overall consistency of our models provides at least a plausible
interpretation of how to connect the baryon cycle to CGM absorption
line observations, which will help develop intuition and testable predictions
for future observational probes of CGM gas.

\section{Acknowledgements}

Support for program GO 11598 was provided by NASA through a grant from
the Space Telescope Science Institute, operated by the Association of
Universities for Research in Astronomy, Inc., under NASA contract NAS
5-26555. Much of the data presented herein were obtained at the W.M. Keck
Observatory, which operates as a scientific partnership among the
California Institute of Technology, the University of California and
the National Aeronautics and Space Administration.  The Observatory
was made possible by the generous financial support of the W.M. Keck
Foundation. The authors wish to recognise and acknowledge the very
significant cultural role and reverence that the summit of Mauna Kea has
always had within the indigenous Hawaiian community. We are most fortunate
to have the opportunity to conduct observations from this mountain.

Additionally, partial support for this work came from NASA ATP NNX12AH86G,
HST grants HST-GO-11598 and HST-GO-12248, NASA ADP grant NNX08AJ44G, NSF
grants AST-0847667 and AST-133514, and the South African National Research
Foundation's Research Chairs program.  The simulations used here were run
on computing facilities owned by the Carnegie Observatories. Computing
resources used for this work were made possible by a grant from the the
Ahmanson foundation, and through grant DMS-0619881 from the National
Science Foundation.
{\it Facilities:} Keck: I (HIRES), {\it Hubble} (COS)

\clearpage

\end{document}